\documentclass[12pt]{article}
\pdfoutput=1
\usepackage{myjheppub}
\usepackage{graphicx,booktabs,latexsym,amssymb,amsmath,bm,epsfig,psfrag,float}
\hypersetup{pdfnewwindow=true}
\preprint{
\begin{flushright}	

	TUM-HEP-1064/16,\\
	IPPP/16/101
\end{flushright}}

\title{NNLL resummation for the associated production of a top pair and a Higgs boson at the LHC}

\author[a,]{Alessandro Broggio,}
\author[b,c]{Andrea Ferroglia,}
\author[d]{Ben D. Pecjak,}
\author[e,f,g]{and Li Lin Yang}

\affiliation[a]{Physik Department T31, Technische Universit\"at M\"unchen,
James Franck-Stra{\ss}e 1, D-85748 Garching, Germany}
\emailAdd{alessandro.broggio@tum.de}

\affiliation[b]{Physics Department, New York City College of Technology, The City
	University of New York, 
	300 Jay Street, Brooklyn, NY 11201 USA}

\affiliation[c]{The Graduate School and University Center,
The City University of New York, 365 Fifth Avenue,
New York, NY 10016  USA}
\emailAdd{aferroglia@citytech.cuny.edu}

\affiliation[d]{Institute for Particle Physics Phenomenology, Ogden Centre for Fundamental Physics,
	Department of Physics, University of Durham, Science Laboratories,
	South Rd, Durham DH1 3LE, United Kingdom}
\emailAdd{ben.pecjak@durham.ac.uk}

\affiliation[e]{School of Physics and State Key Laboratory of Nuclear Physics and Technology,
Peking University, Beijing 100871, China}
\affiliation[f]{Collaborative Innovation Center of Quantum Matter, Beijing, China}
\affiliation[g]{Center for High Energy Physics, Peking University, Beijing 100871, China}
\emailAdd{yanglilin@pku.edu.cn}

\abstract{We study the resummation of soft gluon emission corrections
  to the production of a top-antitop pair in association with a Higgs
  boson at the Large Hadron Collider. Starting from a soft-gluon
  resummation formula derived in previous work, we develop a bespoke
  parton-level Monte Carlo program which can be used to calculate the
  total cross section along with differential distributions.  We use
  this tool to study the phenomenological impact of the resummation to
  next-to-next-to-leading logarithmic (NNLL) accuracy, finding that
  these corrections increase the total cross section and the
  differential distributions with respect to NLO calculations of the
  same observables.}

\begin{document}
\newcommand{\Green}[1]{\textcolor{green}{#1}}
\newcommand{\Blue}[1]{\textcolor{blue}{#1}}
\newcommand{\Cyan}[1]{\textcolor{cyan}{#1}}
\newcommand{\Magenta}[1]{\textcolor{magenta}{#1}}
\newcommand{\alert}[1]{{\bf \color{red}[{#1}]}}

\newcommand{\be}{\begin{equation}}
\newcommand{\ee}{\end{equation}}

\newcommand{\nn}{\nonumber}
\def\ff{f\hspace{-0.3cm}f}
\def\mgamc{{{\tt \small MG5\_aMC}}}

%\vspace*{-1.5cm}
\maketitle

%\newpage

\section{Introduction}
\label{sec:intro}

The associated production of a top-quark pair and a Higgs boson can
provide direct information on the Yukawa coupling of the Higgs boson
to the top quark, which is crucial for verifying the origin of fermion
masses and may shed light on the hierarchy problem related to the mass
of the Higgs boson.  For this reason, experimental collaborations at
the Large Hadron Collider (LHC) are actively searching for this
Higgs-boson production mode in the currently ongoing Run II. The
Standard Model (SM) cross section for this process at a center-of-mass
energy of $13$~TeV is quite small, of the order of $0.5$~pb.

Differences between the measured cross section and the corresponding
SM predictions could indicate the presence of new physics which
modifies the top-quark Yukawa coupling. Consequently, a large amount
of work has been devoted to the study of this process beyond leading
order (LO) in the SM. The LO cross section scales as $\alpha_s^2
\alpha$, where $\alpha_s$ and $\alpha$ denote the strong coupling
constant and the electromagnetic fine structure constant,
respectively. The next-to-leading order (NLO) QCD corrections to this
process were first evaluated more than ten years ago
\cite{Beenakker:2001rj,Beenakker:2002nc,Reina:2001bc,Reina:2001sf,Dawson:2002tg,Dawson:2003zu}. This
process also served as a benchmark for validating automated tools for
NLO calculations; in \cite{Frederix:2011zi,Garzelli:2011vp} the NLO
corrections were calculated automatically and interfaced with Monte
Carlo event generators, thus including parton shower and hadronization 
effects. Electroweak corrections to this process were studied in
\cite{Yu:2014cka,Frixione:2014qaa,Frixione:2015zaa}. NLO QCD and
electroweak corrections were included in the {\tt{POWHEG}} framework
in \cite{Hartanto:2015uka}. In \cite{Denner:2015yca} the NLO
corrections to the associated production of a top pair and a Higgs
boson were studied by considering also the decay of the top quark and
off-shell effects. The cross section for the associated production of a top pair, a Higgs boson and an additional jet at NLO was evaluated in \cite{vanDeurzen:2013xla}.

Perturbative calculations for the $t\bar{t}H$ production process
are difficult and involved, due to the presence of five external legs, four of which carry color
charges. Consequently, it is not likely that the next-to-next-to-leading order (NNLO) QCD
corrections for this process will be computed in the near
future. For this reason, the impact of soft gluon emission corrections beyond 
NLO was the subject of recent studies. In \cite{Kulesza:2015vda} the
soft gluon emission corrections to the total $t \bar{t} H$ cross
section in the production threshold limit were evaluated up to
next-to-leading logarithmic (NLL) accuracy; the production threshold
is defined as the kinematic region in which the partonic
center-of-mass energy approaches $2 m_t + m_H$, which is the minimal
energy of the final state. In \cite{Broggio:2015lya}, on the other
hand, we applied Soft-Collinear Effective Theory (SCET) methods\footnote{See \cite{Becher:2014oda} for an introduction to SCET.}
in order to study the impact of soft-gluon corrections to the associated production
of a top pair and a Higgs boson in the partonic threshold
limit\footnote{Often this limit is referred to as PIM kinematics. The
  acronym PIM stands for Pair Invariant Mass and was extensively
  employed in the context of top-quark pair production. While the
  generalization to our case is trivial, the word ``pair'' should not
  be applied to the process under study here, where the final state
  invariant mass involves 3 particles.}, i.e.  in the limit where the
partonic center-of-mass energy approaches the invariant mass $M$ of
the $t \bar{t} H$ final state. The mass $M$ is bounded from above
only by the hadronic center-of-mass energy. In \cite{Broggio:2015lya}
a resummation formula for the soft emission corrections was derived
and all of the elements necessary for the evaluation of that formula
to next-to-next-to-leading logarithmic (NNLL) accuracy were
evaluated. By using these results, a study of the approximate NNLO
corrections originating from soft gluon emission in the partonic
threshold limit was carried out. In particular, an in-house parton
level Monte Carlo program was developed and employed to evaluate the
total cross section and several differential distributions.  However,
a direct numerical evaluation of the soft gluon emission corrections
to NNLL was not performed in \cite{Broggio:2015lya}.
Recently, results for the total cross section and invariant mass distribution at NLL accuracy in the partonic threshold limit were presented in \cite{Kulesza:2016vnq}.

From the technical point of view, the associated production of a
top pair and a W boson shows several similarities to the associated
production of a top pair and a Higgs boson. However, the former
process involves only one partonic production channel in the partonic
threshold limit, namely the quark annihilation channel, while the
latter also receives large contributions from the gluon fusion
channel.  For this reason some of us recently studied the resummation
of the soft gluon corrections in the partonic threshold limit to $t
\bar{t} W$ production \cite{Broggio:2016zgg}. In that work the
resummation was carried out up to NNLL accuracy in Mellin space. An
in-house parton level Monte Carlo program for the numerical evaluation
of the resummation formulas was developed and employed to obtain
predictions for the total cross section and several differential
distributions at the LHC operating at a center-of-mass energy of $8$ and
$13$~TeV. (The NNLL resummation in the partonic threshold limit for $t
\bar{t} W$ production in momentum space was studied in
\cite{Li:2014ula}.)

By building upon the results  of \cite{Broggio:2015lya} and \cite{Broggio:2016zgg}, in this paper we study the resummation of soft gluon emission corrections to the associated production of a top-quark pair and a Higgs boson in Mellin space. We developed an in-house parton level Monte Carlo code which allows us to evaluate numerically soft emission corrections to this process up to NNLL accuracy. By matching these results with complete NLO calculations carried out with   \verb!MadGraph5_aMC@NLO! \cite{Alwall:2014hca} (which we will indicate with {\mgamc} in the rest of this paper) we obtain predictions for the total cross section and several differential distributions which are valid to NLO+NNLL accuracy.  We also compute the observables at NLO+NLL accuracy
and using NNLO approximations of the NLO+NNLL results, and show that these less precise computations 
miss important effects.

The paper is organized as follows: In Section \ref{sec:outline} we review the salient features of the technique employed to obtain and evaluate the relevant resummation formulas. In Section \ref{sec:numbers} we present predictions, valid to 
NLO+NNLL accuracy, for the total cross section and several differential distributions for the associated production of a top pair and a Higgs boson at the LHC operating at a center-of-mass energy of $13$~TeV. Finally, Section~\ref{sec:conclusions} contains our conclusions.

\section{Outline of the Calculation}
\label{sec:outline}

The associated production of a top quark pair and a Higgs boson receives contributions from the partonic process 
\begin{align}
i(p_1) +  j(p_2) \longrightarrow t(p_3) +\bar{t} (p_4) + H(p_5)  + X \, , 
\label{eq:partproc}
\end{align}
where $ij \in \{q \bar{q}, \bar{q}q, gg\}$ at lowest order in QCD, and
$X$ indicates the unobserved partonic final-state radiation. Two
Mandelstam invariants play a crucial role in our discussion:
\begin{align}
\hat{s} = (p_1+p_2)^2  = 2 p_1 \cdot p_2 \, , \quad \text{and} \quad
M^2 = \left(p_3 + p_4 + p_5 \right)^2 \, .  
\end{align}
The soft or partonic threshold limit is defined as the kinematic situation in which
\begin{align}
z \equiv \frac{M^2}{\hat{s}} \rightarrow 1 \, .
\end{align}
In this region, the unobserved final state can contain only soft radiation.

The factorization formula for the QCD cross section in the partonic 
threshold limit was derived in
\cite{Broggio:2015lya} and reads
\begin{align}
\sigma \left(s,m_t,m_H \right) =& \frac{1}{2 s} \int_{\tau_{\text{min}}}^{1} \!\!\! d \tau \int^1_{\tau} \frac{dz}{\sqrt{z}} \sum_{ij} \ff_{ij} \left( \frac{\tau}{z}, \mu \right)\nonumber \\
& \times \int d\text{PS}_{t\bar{t}H} \mbox{Tr}\left[\mathbf{H}_{ij}\left(\{p\},\mu\right) \mathbf{S}_{ij}\left(\frac{M (1-z)}{\sqrt{z}},\{p\},\mu\right)  \right] \, .
\label{eq:factorization}
\end{align}
In (\ref{eq:factorization}), $s$ indicates the square of the hadronic 
center-of-mass energy and 
\begin{align}
\tau_{\text{min}} = \frac{\left(2 m_t + m_H\right)^2}{s} \, , \qquad \tau = \frac{M^2}{s} \, .
\end{align}
We use the symbol $\{p\}$ to indicate the set of external momenta
$p_1, \cdots, p_5$. The trace $\mbox{Tr}\left[\mathbf{H}_{ij} \mathbf{S}_{ij}
\right]$ is proportional to the spin and color averaged squared matrix
element for $t \bar{t} H + X_s$ production in the process initiated by
the two partons $i$ and $j$, where $X_s$ indicates the unobserved soft
gluons in the final state. 
The hard functions $\mathbf{H}_{ij}$, which are matrices in color space,
are obtained from the color decomposed virtual corrections to the $2
\to 3$ tree-level process.  The soft functions $\mathbf{S}_{ij}$ (which
are also matrices in color space) are related to color-decomposed real
emission corrections in the soft limit; they depend on plus
distributions of the form
\begin{align}
P'_n(z) \equiv \left[\frac{1}{(1-z)}\ln^n\left(\frac{M^2 (1-z)^2}{\mu^2 z}\right)\right]_+ \, ,
\label{eq:plusdist}
\end{align} 
as well as on the Dirac delta function of argument $(1-z)$. 
The parton luminosity functions $\ff_{ij}$ are defined as the
convolutions of the parton distribution functions (PDFs) for the
partons $i$ and $j$ in the protons $N_1$ and $N_2$:
\begin{align}
\ff_{ij} \left(y, \mu\right) = \int_{y}^{1} \frac{dx}{x} f_{i/N_1}\left(x,\mu \right)
f_{j/N_2}\left(\frac{y}{x},\mu \right) \, .
\end{align}
In the soft limit the indices $ij\in\{q\bar{q},\bar{q}q,gg\}$, as
  at LO.  The hard and soft functions are two-by-two matrices for
  $q\bar{q}$-initiated (quark annihilation) processes, and
  three-by-three matrices for $gg$-initiated (gluon fusion) processes.  
Contributions from other production channels such as $\bar{q}g$ and
$qg$ are subleading in the soft limit.  We shall refer to such 
processes collectively as the ``quark-gluon'' or the ``$qg$'' channel in what follows.

The hard functions satisfy renormalization group equations governed by
the soft anomalous dimension matrices $\mathbf{\Gamma}^{ij}_H$, which
depend on the partonic channel considered.  These anomalous dimension
matrices, which are needed to carry out the resummation of soft gluon
corrections, were derived in \cite{Ferroglia:2009ep,Ferroglia:2009ii}.
The hard functions, soft functions, and soft anomalous dimensions must
be computed in fixed-order perturbation theory up to a given order in
$\alpha_s$. In this work we study the resummation up to NNLL
accuracy. For this task we need to evaluate the hard functions, soft
functions and soft anomalous dimensions to NLO. All of these elements
were already evaluated to the order needed here
\cite{Ferroglia:2009ep,Ferroglia:2009ii,Ahrens:2010zv,Broggio:2015lya}. In
particular, the NLO hard functions were evaluated by customizing two
of the one-loop provider programs available on the market, {\tt GoSam}
\cite{Cullen:2011ac,Cullen:2014yla,Binoth:2008uq,Mastrolia:2010nb,Peraro:2014cba}
and {\tt Openloops} \cite{Cascioli:2011va}.  The
numerical evaluation of the hard functions for this work has been
performed by using a modified version of {\tt Openloops}  in combination with {\tt
  Collier}
\cite{Denner:2002ii,Denner:2005nn,Denner:2010tr,Denner:2014gla,Denner:2016kdg}.
{\tt GoSam} in combination with {\tt Ninja}
\cite{Peraro:2014cba,Mastrolia:2012bu,vanDeurzen:2013saa} was used to
cross-check our results.

The resummation formula for the associated production of a $t \bar{t}
H$ final state in Mellin space is similar to the one which was derived
for the production of a $t\bar{t}W$ final state in
\cite{Broggio:2016zgg} and reads
\begin{align}
\sigma(s,m_t,m_H) = & \frac{1}{2 s} \int_{\tau_{\text{min}}}^{1} \frac{d \tau}{\tau} \frac{1}{2 \pi i} \int_{c - i \infty}^{c + i \infty} dN \tau^{-N} \sum_{ij} \widetilde{\ff}_{ij}\left(N, \mu \right) \int d \text{PS}_{t \bar{t} H} \, \widetilde{c}_{ij} \left(N,\mu\right) \, ,
\label{eq:Mellinfac}
\end{align}
where we introduced the Mellin transform of the luminosity functions $\widetilde{\ff}_{ij}$, and 
\begin{align}
\widetilde{c}_{ij} \left(N,\mu\right) \equiv \mbox{Tr} \left[\mathbf{H}_{ij} \left( \{p\},\mu\right) \mathbf{\widetilde{s}}_{ij} \left(\ln\frac{M^2}{\bar{N} \mu^2}, \{p\}, \mu\right) \right] \, .
\label{eq:Mellinc}
\end{align}
Since the soft limit $z \to 1$ corresponds to the limit $N \to \infty$
in Mellin space, we neglected terms suppressed by powers of $1/N$ in
(\ref{eq:Mellinfac}). Furthermore, in (\ref{eq:Mellinc}) we employed
the notation $\bar{N} = N e^{\gamma_E}$. The function
$\mathbf{\widetilde{s}}_{ij}$ is the Mellin transform of the soft function
$\mathbf{S}_{ij}$ found in (\ref{eq:factorization}).

The hard and soft functions in (\ref{eq:Mellinfac}) can be evaluated
in fixed order perturbation theory at scales at which they are free
from large logarithms. We indicate these scales with $\mu_h$ and
$\mu_s$, respectively. Subsequently, by solving the renormalization group (RG) equations for
the hard and soft functions one can evolve the hard scattering kernels
in (\ref{eq:Mellinc}) to the factorization scale $\mu_f$. One obtains
\begin{align}
\widetilde{c}_{ij}(N,\mu_f) =  
\mbox{Tr} \Bigg[&\widetilde{\mathbf{U}}_{ij}(\!\bar{N},\{p\},\mu_f,\mu_h,\mu_s) \, \mathbf{H}_{ij}( \{p\},\mu_h) \, \widetilde{\mathbf{U}}_{ij}^{\dagger}(\!\bar{N},\{p\},\mu_f,\mu_h,\mu_s)
\nn \\
& \times \widetilde{\mathbf{s}}_{ij}\left(\ln\frac{M^2}{\bar{N}^2 \mu_s^2},\{p\},\mu_s\right)\Bigg] \,  .
\label{eq:Mellinresum}
\end{align}
Large logarithmic corrections depending on the ratio
of the scales $\mu_h$ and $\mu_s$ are resummed in the
channel-dependent matrix-valued evolution factors
$\widetilde{\mathbf{U}}$. The expression for the evolution factors is
\begin{align}
\widetilde{\mathbf{U}} \left(\bar{N},\{p\},\mu_f, \mu_h, \mu_s \right) =& \exp \Biggl\{
2 S_{\Gamma_{\text{cusp}}} (\mu_h, \mu_s) - a_{\Gamma_{\text{cusp}}} (\mu_h, \mu_s) \ln 
\frac{M^2}{\mu_h^2} + a_{\Gamma_{\text{cusp}}} (\mu_f, \mu_s) \ln \bar{N}^2 
\nonumber \\
&+ 2 a_{\gamma^\phi}(\mu_s,\mu_f) 
\Biggr\} \times \mathbf{u} \left( \{p\},\mu_h,\mu_s \right) ,
\label{eq:Uevolution}
\end{align}
which is formally identical to the expression found for the
corresponding quantity in carrying out the resummation for $t\bar{t}
W$ production. For the definition of the various RG factors 
appearing in (\ref{eq:Uevolution}) we refer the reader to
\cite{Broggio:2016zgg}.  However, while for $t\bar{t}W$ production one
needs to consider the evolution factor in the quark-annihilation
channel only, for $t \bar{t} H$ production one also needs to evaluate
the appropriate anomalous dimensions and evolution factor for the
gluon fusion channel. 

The functions $\mathbf{U}$ in (\ref{eq:Uevolution}) depend on $\alpha_s$
evaluated at three different scales: $\mu_h$, $\mu_s$ and $\mu_f$.  In
practice, it is convenient to rewrite the evolution factors in terms
of $\alpha_s(\mu_h)$ only. This can be done by employing the running
of $\alpha_s$ at three loops \cite{Moch:2005ba}. By doing this, 
logarithms such as $\ln(\mu_h/\mu_s)$ appear explicitly in the formula
for the evolution matrix, which becomes \cite{Broggio:2016zgg}
\begin{align}
\widetilde{\mathbf{U}}\left(\bar{N}, \{p\},\mu_f, \mu_h, \mu_s \right) =& \exp \Biggl\{
\frac{4 \pi}{\alpha_s(\mu_h)} g_1 \left( \lambda, \lambda_f \right) +  g_2 \left( \lambda, \lambda_f \right) + \frac{\alpha_s(\mu_h)}{4 \pi}  g_3 \left( \lambda, \lambda_f \right) + \cdots \Biggr\} \nn \\
& \times \mathbf{u} ( \{p\},\mu_h, \mu_s)\, ,
\label{eq:Uevolutiongs}
\end{align} 
with
\begin{align}
\lambda = \frac{\alpha_s(\mu_h)}{2 \pi} \beta_0 \ln{\frac{\mu_h}{\mu_s}} \, ,
\qquad
\lambda_f = \frac{\alpha_s(\mu_h)}{2 \pi} \beta_0 \ln{\frac{\mu_h}{\mu_f}}\, .
\end{align}
The leading logarithmic (LL) function $g_1$, the NLL function $g_2$, and the NNLL function $g_3$ can be
obtained starting from (\ref{eq:Uevolution}). One can see that the
l.h.s of (\ref{eq:Mellinresum}) is independent of $\mu_h$ and $\mu_s$
if the evolution factors and the hard and soft functions are known to
all orders in perturbation theory. This is impossible in practice, which
introduces a residual dependence on the choice of the scales $\mu_h$
and $\mu_s$ in any numerical evaluation of (\ref{eq:Uevolution}) or
(\ref{eq:Uevolutiongs}). The hard and soft functions are free from
large logarithms if one chooses $\mu_h \sim M$ and $\mu_s \sim
M/\bar{N}$. It is well known that one then faces the presence of a
branch cut for large values of $N$ in the hard scattering kernel,
whose existence is related to the Landau pole in $\alpha_s$. In this
work, we choose the integration path in the complex $N$ plane when
evaluating the inverse Mellin transform according to the
{\emph{Minimal Prescription}} (MP) introduced in \cite{Catani:1996yz}.
In the numerics, we need the parton luminosity functions in Mellin
space. These can be constructed using techniques described in 
\cite{Bonvini:2012sh, Bonvini:2014joa}.

\section{Numerical Results}
\label{sec:numbers}

\begin{table}[t]
	\begin{center}
		\def\arraystretch{1.3}
		\begin{tabular}{|c|c||c|c|}
			\hline $M_W$ & $80.419$~GeV & $m_t$ & $173.2$~GeV\\ 
			\hline $M_Z$ &  $91.1876$~GeV & $m_H$ & $125$~GeV \\ 
			\hline $G_F$ & $1.16639 \times 10^{-5}$~GeV$^{-2}$ & $\alpha_s \left(M_Z\right)$ & from MMHT 2014 PDFs \\ 
			\hline 
		\end{tabular} 
		\caption{Input parameters employed throughout the calculation. \label{tab:tabGmu}}
	\end{center}
\end{table}

In this section we present predictions for the total cross section and
differential distributions for the $t \bar{t} H$ production
process. The main goal of this work is to obtain predictions for
physical observables which are valid to NLO+NNLL accuracy. However, we
also perform some systematic studies meant to provide insight into the
validity of various approximations to this state-of-the-art result.
In all cases, we use the input parameters listed in
Table~\ref{tab:tabGmu}, and MMHT 2014 PDFs
\cite{Harland-Lang:2014zoa}.  We switch PDF orders as appropriate for
a given perturbative approximation according to the scheme given in 
Table~\ref{tab:CSHp13hM}, where we also specify the computer code used in 
each case.

As a preliminary step we check that with our choice of scales and
input parameters the NLO expansion of the NNLL resummation formula
(which we refer to as ``approximate NLO'') provides a satisfactory
approximation to the exact NLO calculation. Such an approximation of
  (\ref{eq:Mellinresum}) captures the leading terms in the Mellin-space
  soft limit ($N\to \infty$) of the NLO cross section, namely the  single
  and double powers of $\ln N$ as well as  $N$-independent
  terms. Even though the $N$-independent terms depend on the Mandelstam 
variables, we will refer to them as ``constant'' terms in what follows.  
Analogous comparisons of approximate NLO and complete NLO calculations
were carried out for $t\bar{t}W$ production in
\cite{Broggio:2016zgg}. In \cite{Broggio:2015lya}, similar comparisons
were also performed for $t\bar{t}H$ production, but with two
differences with respect  to the current work: the renormalization and
factorization scales were fixed (independent of $M$) instead of
dynamic (correlated with $M$), and the leading terms were represented
in momentum space instead of Mellin space.

The NLO approximation mentioned above is easily obtained by setting
$\mu_s=\mu_h=\mu_f$ in the NNLL resummation formula
(\ref{eq:Mellinresum}).  For this reason, the matched NLO+NNLL cross
section is given by
\begin{align}
\sigma^{{\text{NLO+NNLL}}}  =&  \sigma^{{\text{NLO}}}
+\left[ \sigma^{\text{NNLL}}- \sigma^{\text{approx. NLO}}\right]\,.
\label{eq:NLOpNNLLmatching}
\end{align}
The difference of terms in the square brackets contributes at NNLO and
beyond, adding NNLL resummation onto the NLO result.  In
order to study the convergence of resummed perturbation theory, we
will also calculate NLO+NLL results, defined as
\begin{align}
\sigma^{{\text{NLO+NLL}}}  =& 
\sigma^{{\text{NLO}}}  + 
\left[\sigma^{{\text{NLL}}}  -  \sigma^{{\text{NLL expanded to NLO}}} \right]  \, .
\label{eq:NLOpNLLmatching_old}
\end{align}
The difference of terms in the square brackets contributes at NNLO and
beyond, adding NLL resummation onto the NLO result.  However, in
contrast to the approximate NLO result, the constant 
piece of the NLO expansion of the NLL resummation formula contains
explicit dependence on the matching scales $\mu_h$ and $\mu_s$, in
addition to that on $\mu_f$. The numerical dependence on these
scales is formally of NNLL order (and is indeed canceled through
$\mu_s$ and $\mu_h$ dependence in the NLO hard and soft functions in the
NNLL result), and provides an additional handle on estimating the size
of NNLL corrections using the NLL resummation formula.

While we are mainly interested in NNLL resummation effects, it is also
interesting to study to what extent these all-orders corrections are
approximated by their NNLO truncation.  To this end, we consider
``approximate NNLO'' calculations based on the NNLL resummation
formula (\ref{eq:Mellinresum}). Approximate NNLO calculations include
all powers of $\ln N$ and part of the constant terms from a complete
NNLO calculation, but neglect terms which vanish as $N\to
\infty$. Since the constant terms are not fully determined by an NNLL
calculation (only their $\mu$-dependence is, through the RG
equations), there is some freedom as to how to construct such
approximations.  

Here we consider two possibilities.  The first follows the procedure
used in \cite{Broggio:2016zgg} for the case of $t \bar{t} W$
production.  A detailed description of which constant pieces are
included in that NNLO approximation can be found in Section 4 of
\cite{Broggio:2016zgg}\footnote{In \cite{Broggio:2015lya} such
  approximate NNLO formulas were obtained starting from the
  resummation formula in momentum space, and thus differ from Mellin
  space results through power corrections and constant terms. However,
  we have checked that the two approaches lead to results which are
  numerically almost identical.}.  We match these NNLO corrections,
obtained in the soft limit, with the NLO ones in the usual way:
\begin{align}
\sigma^{\text{nNLO}} =\sigma^{\text{NLO}} 
+\left[\sigma^{\text{approx. NNLO}} - \sigma^{\text{approx. NLO}}\right] \, ,
\label{eq:NNLOmatching}
\end{align}
where we introduced the acronym nNLO to indicate approximate NNLO
corrections matched to full NLO calculations.  The second NNLO
approximation we consider is based on the direct expansion of the
NLO+NNLL result to NNLO.  This differs from the approximate NNLO
result used above by constant terms, which are formally of N$^3$LL
order.  We define this approximation through the matching equation
\begin{align}
\label{eq:NNLLexpanded}
\left(\sigma^{\text{NLO+NNLL}}\right)_{\text{NNLO exp.}} &=
\sigma^{\text{NLO}} +
\left[\sigma^{\text{NNLL expanded to NNLO}}
- \sigma^{\text{approx. NLO}}\right] \,.
\end{align}
In both cases above, the difference of terms in the square brackets is
a pure NNLO correction. Contrary to the approximate NNLO result used
in (\ref{eq:NNLOmatching}), which depends only on $\mu_f$ by
construction, the constant pieces of the NNLO expansion of the NNLL
result in (\ref{eq:NNLLexpanded}) contain explicit dependence on
$\mu_h$ and $\mu_s$, in addition to that on $\mu_f$.  This scale dependence is
formally of N$^3$LL order, and can be used to estimate the size of
such corrections to the NNLL results.  Moreover, the NNLO
approximation (\ref{eq:NNLLexpanded}) differs from the NLO+NNLL result through terms of
N$^3$LO and higher, so comparing the two results gives a direct
measure of how important such terms are numerically. In fact, were
an exact NNLO calculation to appear, adding to it these beyond NNLO
terms would achieve NNLO+NNLL resummation.

\subsection{Scale choices}
\begin{figure}[tp]
	\begin{center}
			\includegraphics[width=10.5cm]{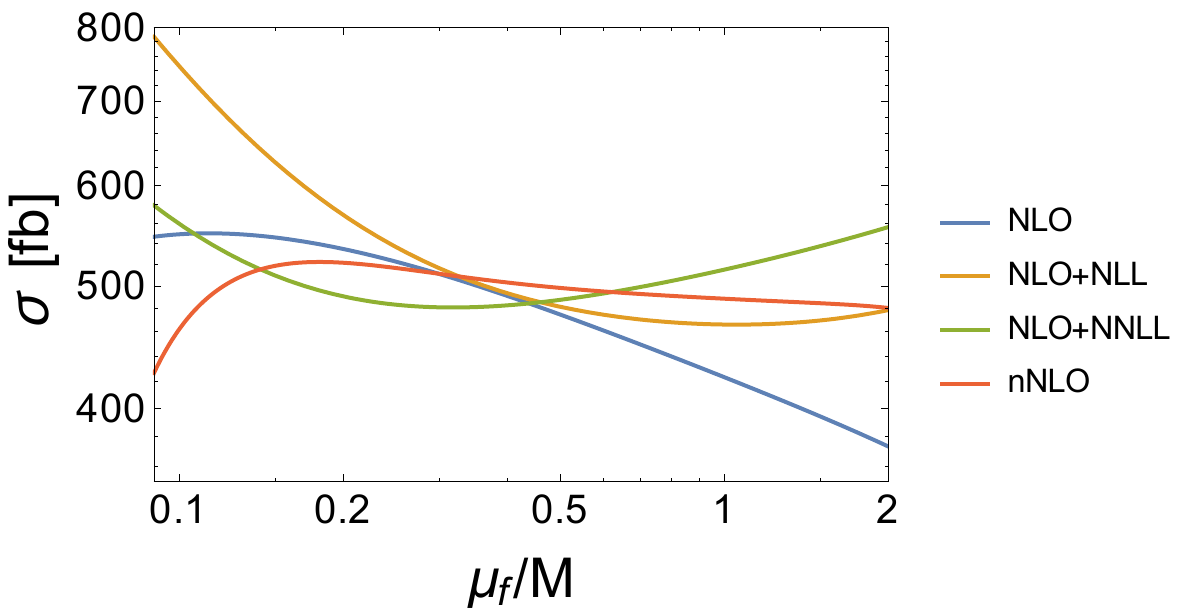} 
	\end{center}
	\caption{Factorization-scale dependence of the total $t\bar{t}H$ production cross section 
	at the LHC with $\sqrt{s}=13$~TeV. The NLO and NLO+NLL curves are obtained 
	using MMHT 2014 NLO PDFs, while the NLO+NNLL and nNLO curves are obtained using  MMHT 2014 NNLO PDFs.
		\label{fig:scaledep}
	}
\end{figure}

Numerical evaluations of the resummed formulas have a residual
dependence on the choice of the hard and soft scales $\mu_h$ and
$\mu_s$. This feature arises from the fact that the various factors in
(\ref{eq:Mellinresum}) have to be evaluated at a given order in
perturbation theory. When the resummation is carried out in Mellin
space the standard default choice of these scales is $\mu_{h,0} = M$ and $\mu_{s,0} =
M/ \bar{N}$ \cite{Ferroglia:2015ivv,Pecjak:2016nee,Broggio:2016zgg}. This choice
is the same one followed in the ``direct QCD'' resummation method
\cite{Catani:1996yz, Bonvini:2012az,Bonvini:2014qga}, and is the one we shall use here.

Furthermore, both the fixed-order and resummed results have a residual
dependence on the factorization scale $\mu_f$. The factorization scale
should be chosen in such a way that logarithms of the ratio $\mu_f/M$
are not large \cite{Collins:1989gx}. Since we are working in the
partonic threshold limit it is natural to choose a dynamical 
value for the factorization scale which is correlated with the final state
invariant mass $M$. Figure~\ref{fig:scaledep} shows the dependence of
the total cross section calculated within various perturbative approximations
on the choice of the ratio $\mu_f/M$ at the LHC with $\sqrt{s}=13$~TeV. One can
observe that the NLO, NLO+NLL and NLO+NNLL curves intersect each other
in the vicinity of $\mu_f/M = 0.5$, while the three curves have a very
different behavior for small values of $\mu_f$.  In addition, Figure
\ref{fig:scaledep} shows that beyond-NLO corrections are quite
significant for  $\mu_f/M \gg 0.5$, as the NLO result falls rather
steeply away to smaller values in that region, while the other three
curves remain reasonably stable. 

Because of these considerations, in the following we employ two
different default choices for the factorization scale, namely
$\mu_{f,0} = M/2$ and $\mu_{f,0} = M$.  The choice $\mu_{f,0} =
  M/2$ may be advantageous because the lower-order perturbative
  results are larger at lower $\mu_f$, so that the apparent
  convergence of the perturbative series is improved, but other than
  this numerical fact there is no obvious reason to exclude the
  natural hard scale $M$ as a default choice so we study this as well.
  In both cases, the uncertainty
associated to the choice of a default value for the scale is estimated
by varying each scale in the interval $[ \mu_{i,0}/2, 2 \mu_{i,0} ]$
($i \in \{s,f,h\}$). 
The scale uncertainty above the central value of
an observable $O$ (the total cross section, or the value of a
differential cross section in a given bin) is then evaluated by
combining in quadrature the quantities
\begin{align}
\Delta O_i^+= \text{max}\{O \left(\kappa_i=1/2\right),  O \left(\kappa_i=1\right), O \left(\kappa_i=2\right)\} - \bar{O} \, , \label{eq:DeltaO}
\end{align}
for $i=f,h,s$. In (\ref{eq:DeltaO}) $\kappa_{i} = \mu_i/\mu_{i,0}$ and $\bar{O}$ is the value of $O$  evaluated by setting all scales to their default values  ($\kappa_i = 1$ for $i=f,h,s$). The scale uncertainty below the central value can be obtained in the same way by combining in quadrature the quantities $\Delta O_i^-$, defined as in (\ref{eq:DeltaO}) but with ``$\text{max}$'' replaced by ``$\text{min}$''. We use this procedure to obtain the perturbative uncertainties given in all of the tables and 
figures that follow.

\subsection{Total cross section}
\label{sec:CS}

% % % % % % % % %
\begin{table}[t]
	\begin{center}
		\def\arraystretch{1.3}
		\begin{tabular}{|c|c|c|c|}
			\hline  order & PDF order & code & $\sigma$ [fb]\\ 
			\hline LO & LO & \mgamc & $ 378.7^{+120.5}_{-85.2} $ \\
			\hline \hline 
			\hline app. NLO & NLO & in-house MC & $ 473.3^{+0.0}_{-28.6} $ \\
			\hline NLO no $qg$ & NLO & \mgamc & $ 482.1^{+10.9}_{-35.1}$ \\
		         \hline NLO & NLO & \mgamc & $ 474.8^{+47.2}_{-51.9} $ \\
				%\hline nNLO (momentum) & NNLO & in-house MC +\mgamc  & $ $ \\
			\hline \hline NLO+NLL & NLO&  in-house MC  +\mgamc  & $480.1^{+57.7}_{-15.7} $ \\
		\hline NLO+NNLL  & NNLO & in-house MC +\mgamc  & $486.4^{+29.9}_{-24.5} $ \\
		\hline 
		\hline nNLO (Mellin) & NNLO & in-house MC +\mgamc  & $497.9^{+18.5}_{-9.4} $ \\
		\hline 
			(NLO+NNLL)$_{\rm NNLO \, exp.}$  & NNLO & in-house MC +\mgamc  & $482.7^{+10.7}_{-21.1} $ \\	
								\hline 
		\end{tabular} 
		\caption{Total cross section for $t \bar{t} H$ production at the
                  LHC with $\sqrt{s} = 13$~TeV and MMHT 2014 PDFs. The default value
                    of the factorization scale is $\mu_{f,0}=M/2$, and
                    the uncertainties are estimated through scale
                    variations of this (and the resummation scales
                    $\mu_s$ and $\mu_h$ when applicable) as explained in the text, see the discussion around (\ref{eq:DeltaO}).
\label{tab:CSHp13hM}}
	\end{center}
\end{table}

\begin{figure}[t]
	\begin{center}
		\includegraphics[width=8.5cm]{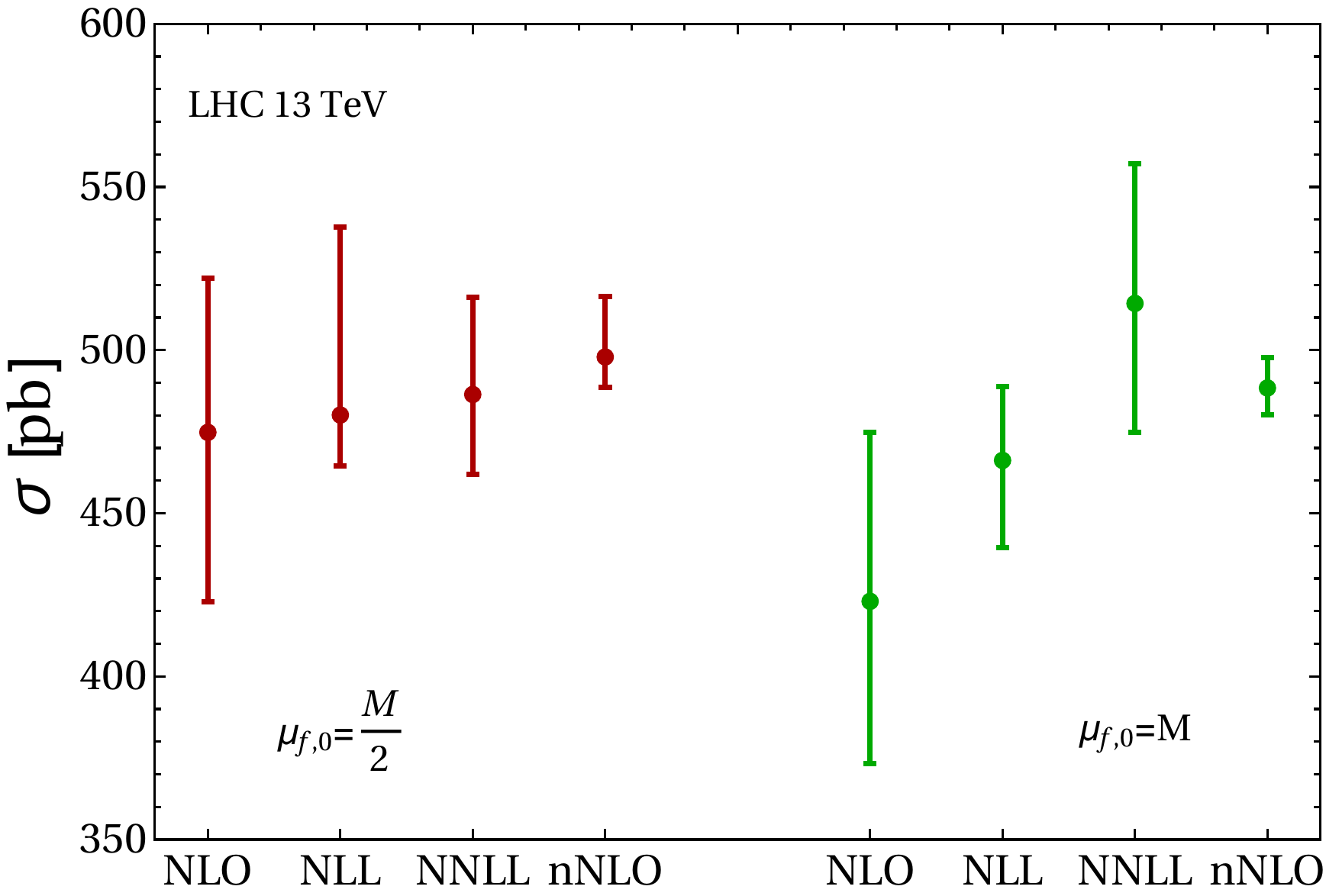} 
	\end{center}
	\caption{Comparison between different perturbative approximations to the total cross section  carried out 
	with the default factorization scale choices $\mu_{f,0}=M/2$ (left) and $\mu_{f,0}=M$ (right).
		The labels ``NLL'' and ``NNLL'' on the horizontal axis indicate NLO+NLL and NLO+NNLL calculations.
		\label{fig:TOTCScomp}
	}
\end{figure}

We begin our analysis by considering the total cross section for the
associated production of a top pair and a Higgs boson at the LHC
operating at a center-of-mass energy of 13~TeV. The results obtained
are summarized in Table~\ref{tab:CSHp13hM}, where we set $\mu_{f,0} =
M/2$, in Table~\ref{tab:CSHp13}, where we set $\mu_{f,0} = M$, and in
Figure~\ref{fig:TOTCScomp}, which presents a visual comparison 
between the main results at the two different scales.

We first compare the approximate NLO corrections generated from NNLL
soft-gluon resummation (second row of each table), with the full NLO
corrections without (third row of each table) and with (fourth row of each
table) the $qg$ channel.  Since the approximate NLO results include
only the leading-power contributions from the gluon fusion and
quark-annihilation channels in the soft limit, the difference between
these results and the NLO corrections without the $qg$ channel gives a
measure of the importance of power corrections away from this
limit. The two results are seen to differ by no more than a few
percent, even though the NLO corrections are large.  This shows that
at NLO the power corrections away from the soft limit for these
channels are quite small.  Comparing the NLO results with and without
the $qg$ channel reveals that this channel contributes 
significantly to the scale uncertainty, in particular when one chooses $\mu_{f,0} = M/2$.
 The fact that
the leading terms in the soft limit make up the bulk of the NLO
correction provides a strong motivation to resum them to all orders.
No information is lost by doing this, as both sources of power
corrections are taken into account by matching with NLO as discussed
above.  Since the power corrections are treated in fixed order, the
perturbative uncertainties associated with them are estimated through
the standard approach of scale variations.

% % % % % % % % %
\begin{table}[t]
	\begin{center}
		\def\arraystretch{1.3}
		\begin{tabular}{|c|c|c|c|}
			\hline  order & PDF order & code & $\sigma$ [fb]\\ 
			\hline LO & LO & \mgamc & $293.5^{+85.2}_{-61.7} $ \\
			\hline \hline
				 app. NLO & NLO & in-house MC & $ 444.7^{+28.6}_{-39.2} $ \\	
			\hline NLO no $qg$ & NLO & \mgamc & $ 447.0^{+35.1}_{-40.4}$ \\
	            	\hline NLO & NLO & \mgamc & $423.0^{+51.9}_{-49.7} $ \\
			\hline \hline NLO+NLL & NLO&  in-house MC  +\mgamc  & $466.2^{+22.9}_{-26.8} $ \\
			\hline NLO+NNLL  & NNLO & in-house MC +\mgamc  & $514.3^{+42.9}_{-39.5} $ \\	
			\hline \hline 			
			 nNLO (Mellin) & NNLO & in-house MC +\mgamc  & $488.4^{+9.4}_{-8.3} $ \\
			\hline (NLO+NNLL)$_{\rm NNLO \, exp.}$  & NNLO & in-house MC +\mgamc  & $485.7^{+6.8}_{-15.0} $ \\
			 \hline 
		\end{tabular} 
		\caption{Total cross section for $t \bar{t} H$ at the LHC with $\sqrt{s} = 13$~TeV and MMHT 2014 PDFs. The results are obtained as in 
Table~\ref{tab:CSHp13hM}, but with the default value of the factorization scale 
chosen instead as $\mu_{f,0}=M$.			\label{tab:CSHp13}}	
	\end{center}
\end{table}
% % % % % % % % % % % % % %

We next turn to the NLO+NLL and NLO+NNLL cross sections, which are the main results of this section.
The exact numbers are given in Tables~\ref{tab:CSHp13hM} and~\ref{tab:CSHp13},  
and a pictorial representation is given in Figure~\ref{fig:TOTCScomp}.
The results for the default scale choice $\mu_{f,0}=M/2$ converge quite nicely.  The 
scale uncertainties get progressively smaller when moving from NLO to NLO+NLL 
to NLO+NNLL, and the higher-order results are roughly within the range predicted
by the uncertainty bands of the lower-order ones.  For $\mu_{f,0}=M$ the 
convergence is still reasonable but not quite as good, mainly because the 
NLO and NLO+NLL  results are noticeably smaller than at $\mu_{f,0}=M/2$.  
Interestingly, the NLO+NLL result  has a smaller scale uncertainty than
the NLO+NNLL one for $\mu_{f,0}=M$, a fact which looks rather accidental 
considering its wider
variation over a larger range of $\mu_f$, as  seen  in Figure~\ref{fig:scaledep}.
However, one should remember that the scales $\mu_h$ and $\mu_s$ are kept fixed at their default values in the NLO+NLL and NLO+NNLL curves of Figure~\ref{fig:scaledep}, while they are varied as explained above in order to obtain the scale uncertainty reported in the tables.

Finally, we discuss the NNLO approximations to the NNLL resummation formula.
The results in Table~\ref{tab:CSHp13hM} show that for $\mu_{f,0}=M/2$ 
the importance of resummation effects beyond NNLO is rather small, roughly at or below 
the 5\% level after taking scale uncertainties into account.
An examination of Table~\ref{tab:CSHp13}
shows that the effects are noticeably  larger at $\mu_{f,0}=M$, approximately at 
the 10\% level. In either case, 
Figure~\ref{fig:TOTCScomp} shows very clearly that the nNLO results 
display an artificially small scale dependence compared to the NLO+NNLL
results, confirming the cautionary statements made in \cite{Broggio:2015lya} about
the reliability of the nNLO scale dependence in estimating higher-order 
perturbative corrections.

 The results in this section highlight the importance of an NNLL 
 calculation.  Taken as a whole, they show that both NLO+NLL
 and approximate NNLO calculations are a poor proxy for the more
 complete NLO+NNLL calculation.  We have considered two default
 scale choices, $\mu_{f,0}=M/2$ and $\mu_{f,0}=M$.  
 However, we should emphasize that  in the end the default scale choice is 
arbitrary, and it would not be unreasonable to combine the envelope
of results from the two choices into a single, larger perturbative uncertainty.
The NLO+NNLL results quoted at either scale would not change significantly
through such a combination.

% % % % % % % % % % % % % % % % % % % % % % % % % % % % % % %

\subsection{Differential distributions}

\begin{figure}[tp]
	\begin{center}
		\begin{tabular}{cc}
			\includegraphics[width=7.2cm]{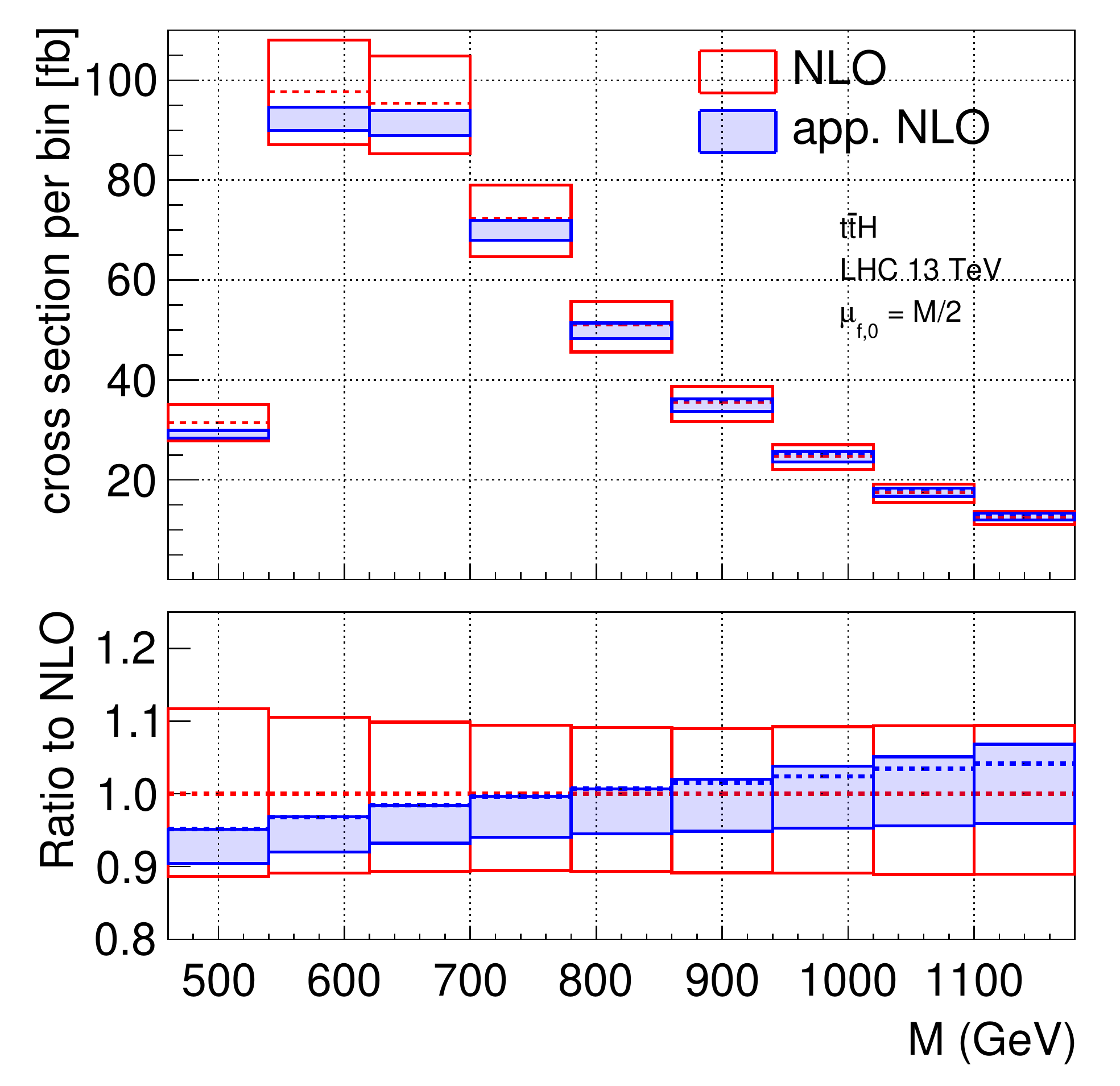} & \includegraphics[width=7.2cm]{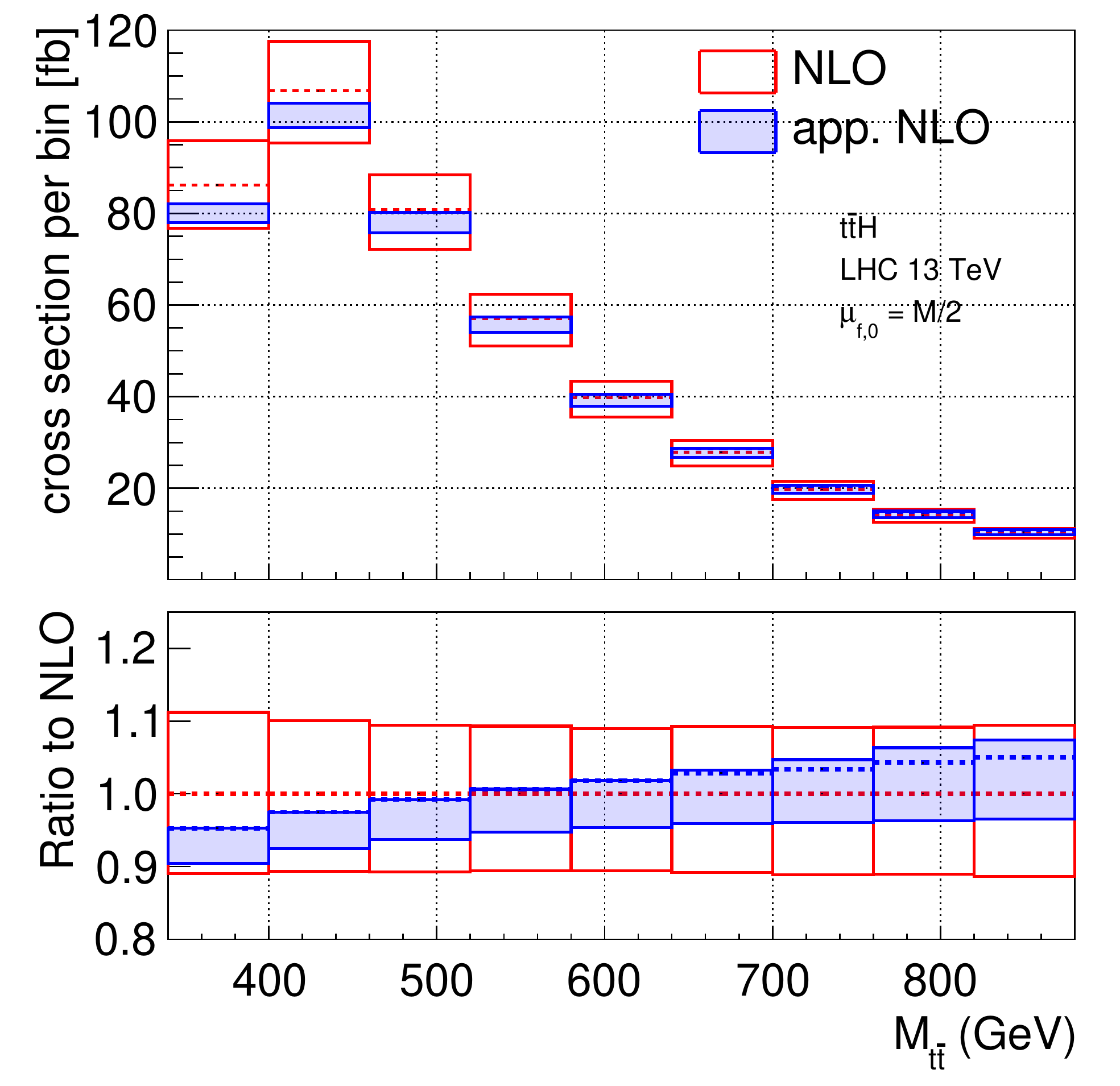} \\
			\includegraphics[width=7.2cm]{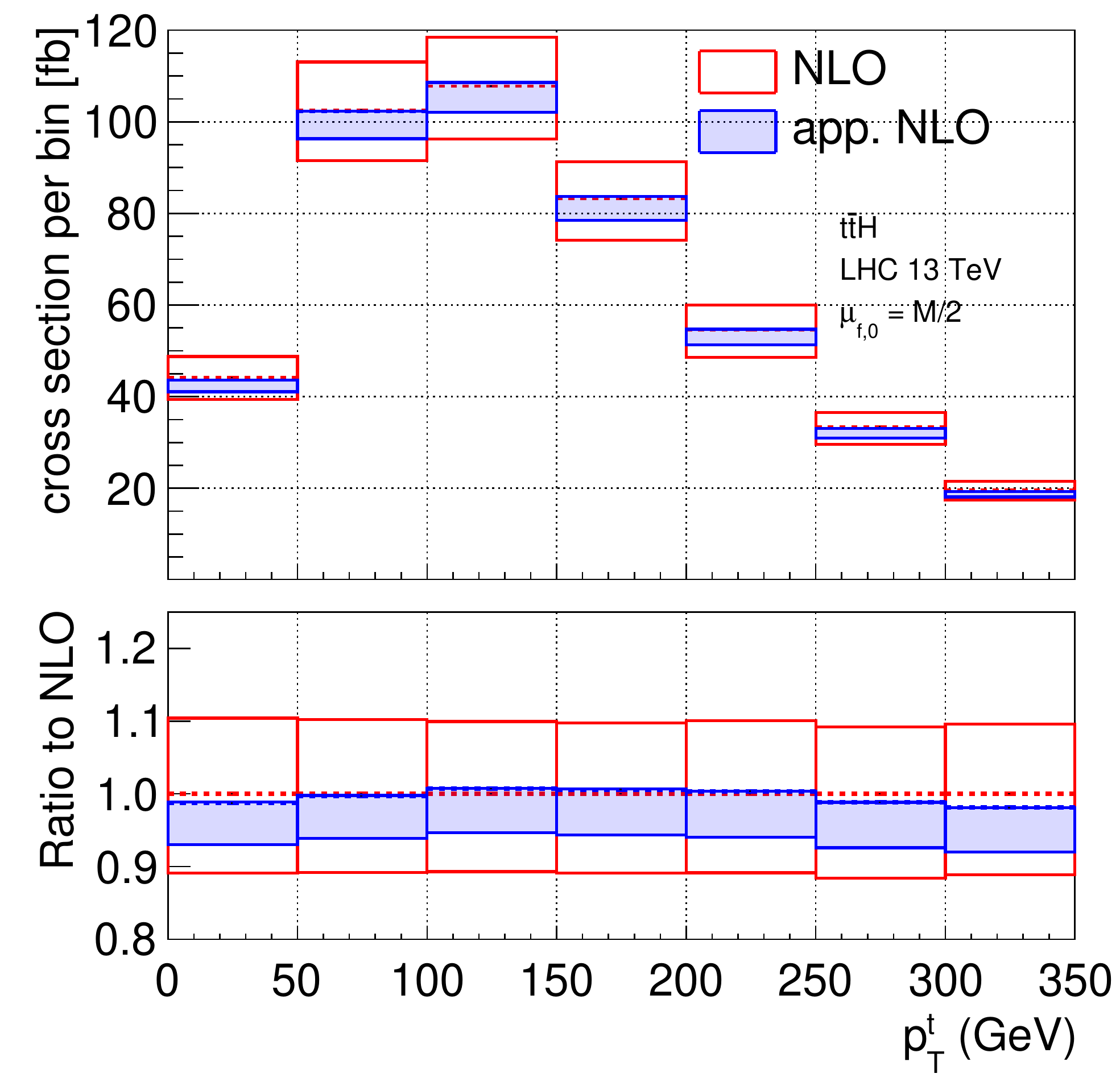} & \includegraphics[width=7.2cm]{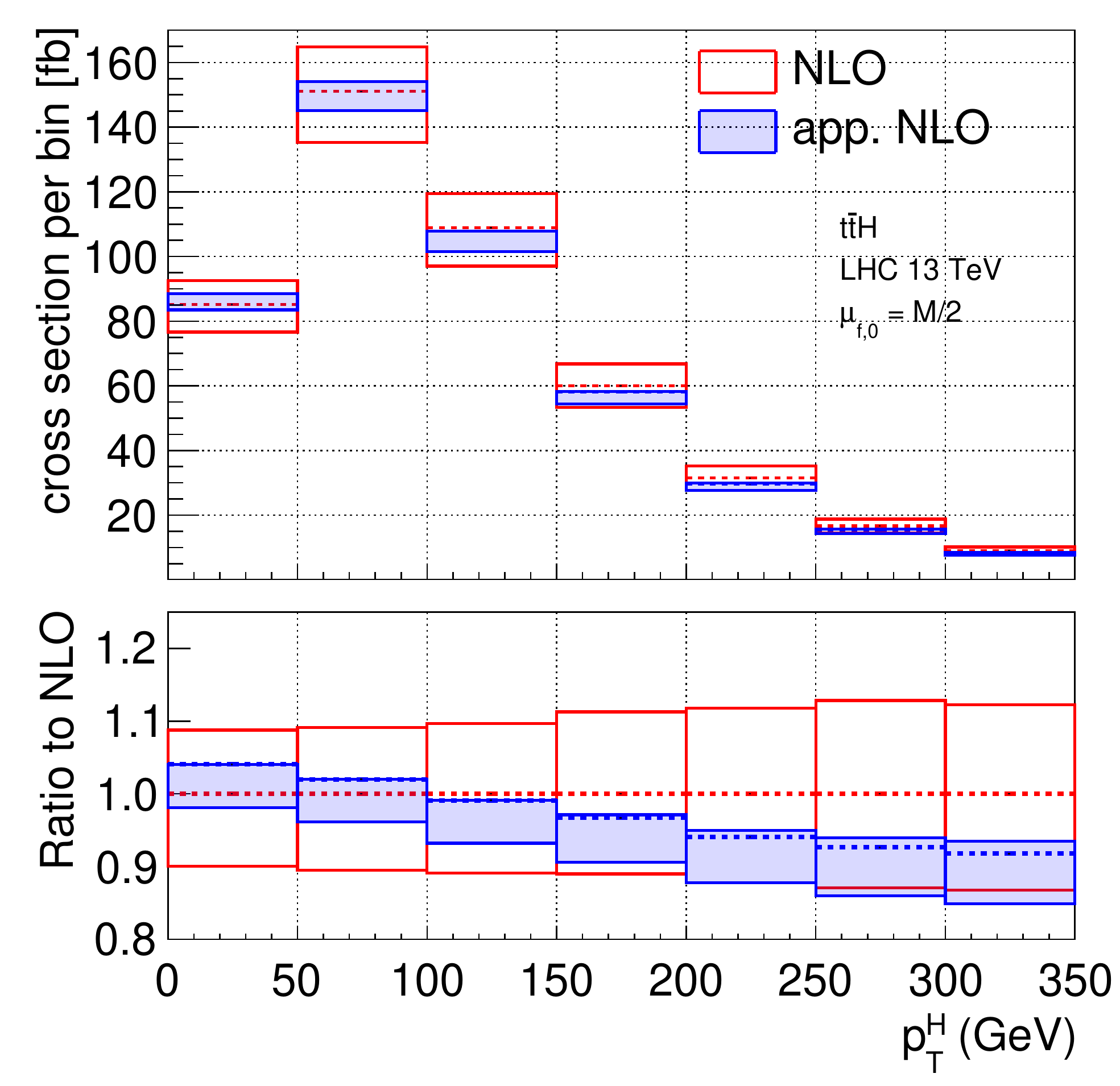} \\
		\end{tabular}
	\end{center}
	\caption{Differential distributions at approximate NLO (blue
          band) compared to the complete NLO (red band). The default
          factorization scale is chosen as $\mu_{f,0}=M/2$, and the
          uncertainty bands are generated through scale variations as
          explained in the text. \label{fig:nLOvsNLOhalfM}}
	
\end{figure}

In this section we discuss results for differential distributions. In
particular, we consider:
\begin{itemize}
	\item the distribution differential with respect to the invariant mass of the top pair and Higgs boson in the final state, $M$;
	\item the distribution differential with respect to the invariant mass of the top-quark pair, $M_{t\bar{t}}$;
	\item the distribution differential with respect to the transverse momentum of the Higgs boson, $p_T^H$;
	\item  the distribution differential with respect to the transverse momentum of the top quark, $p_T^t$.	
\end{itemize}

We first set the default value of the factorization scale to
$\mu_{f,0} = M/2$. Figure~\ref{fig:nLOvsNLOhalfM} shows the comparison
between complete NLO calculations and approximate NLO calculations for
all of the distributions listed above. We observe that for all of the
distributions the approximate NLO scale uncertainty band (in blue) is
included in the NLO scale uncertainty band (bins with the red
frame). However, the approximate NLO uncertainty is smaller than the
NLO uncertainty in all bins. Furthermore the bin-by-bin ratio of the
two distributions, found at the bottom of each panel, shows that the
NLO and approximate NLO corrections have somewhat different shapes.

\begin{figure}[tp]
	\begin{center}
		\begin{tabular}{cc}
			\includegraphics[width=7.2cm]{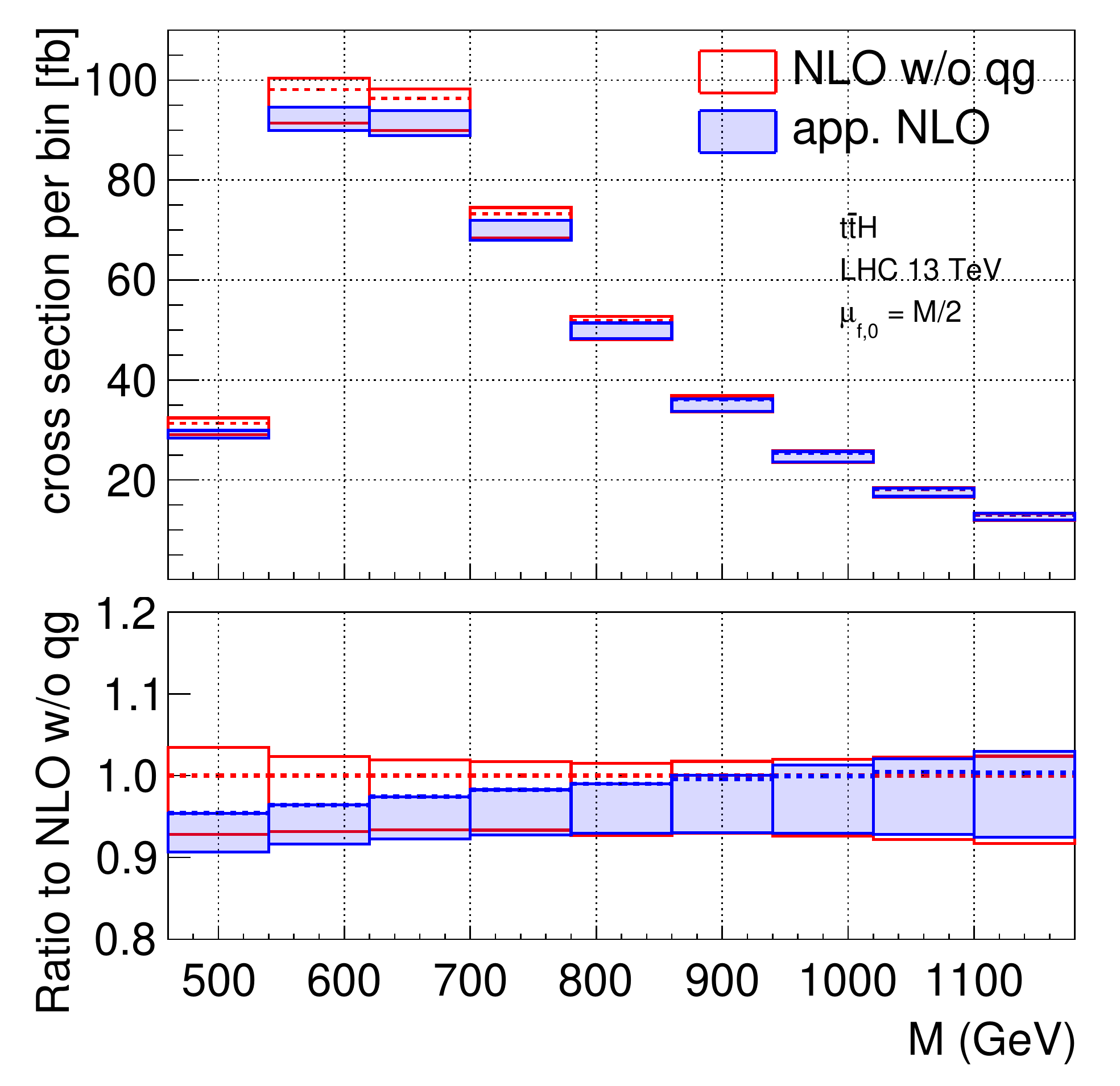} & \includegraphics[width=7.2cm]{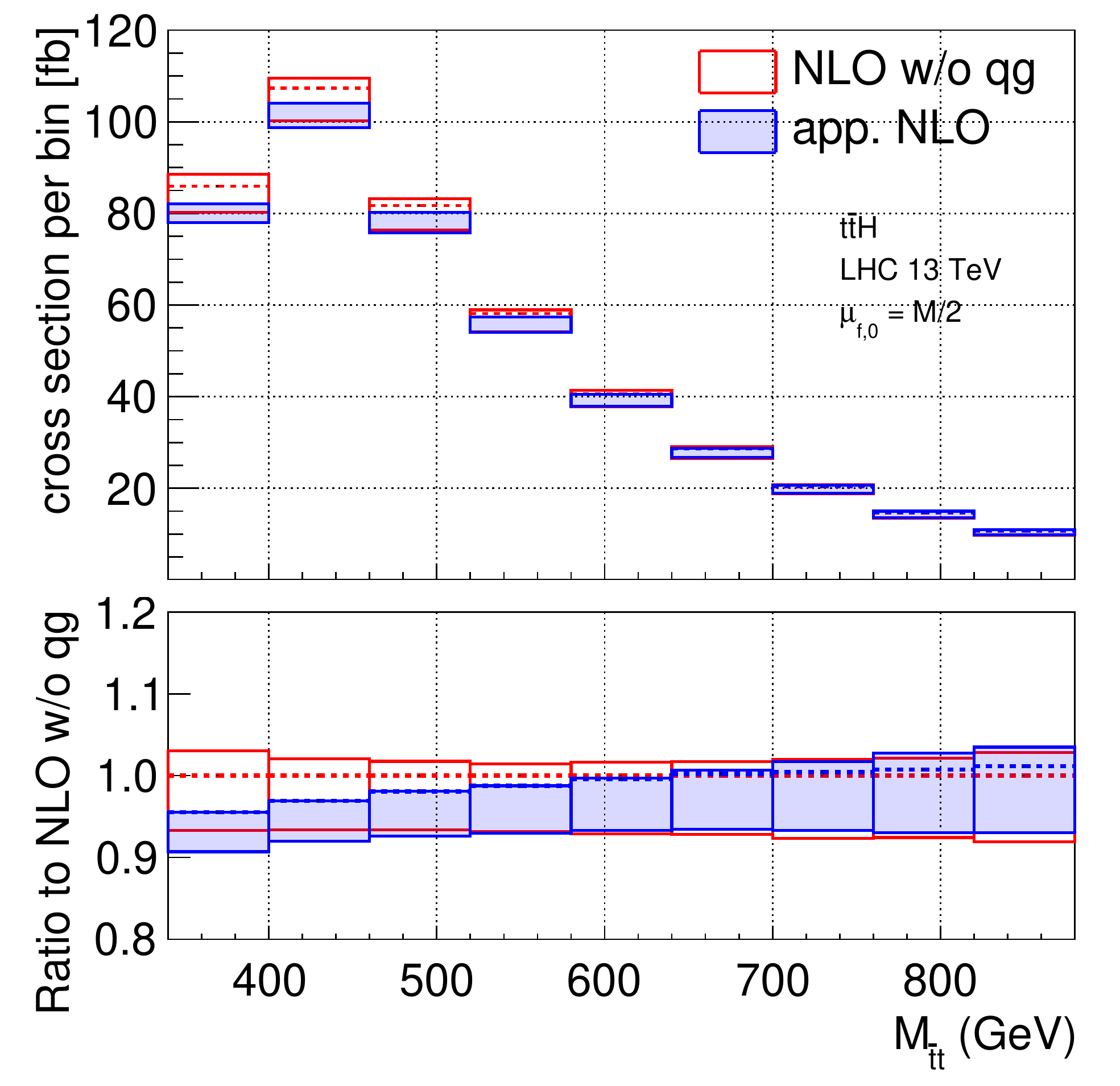} \\
			\includegraphics[width=7.2cm]{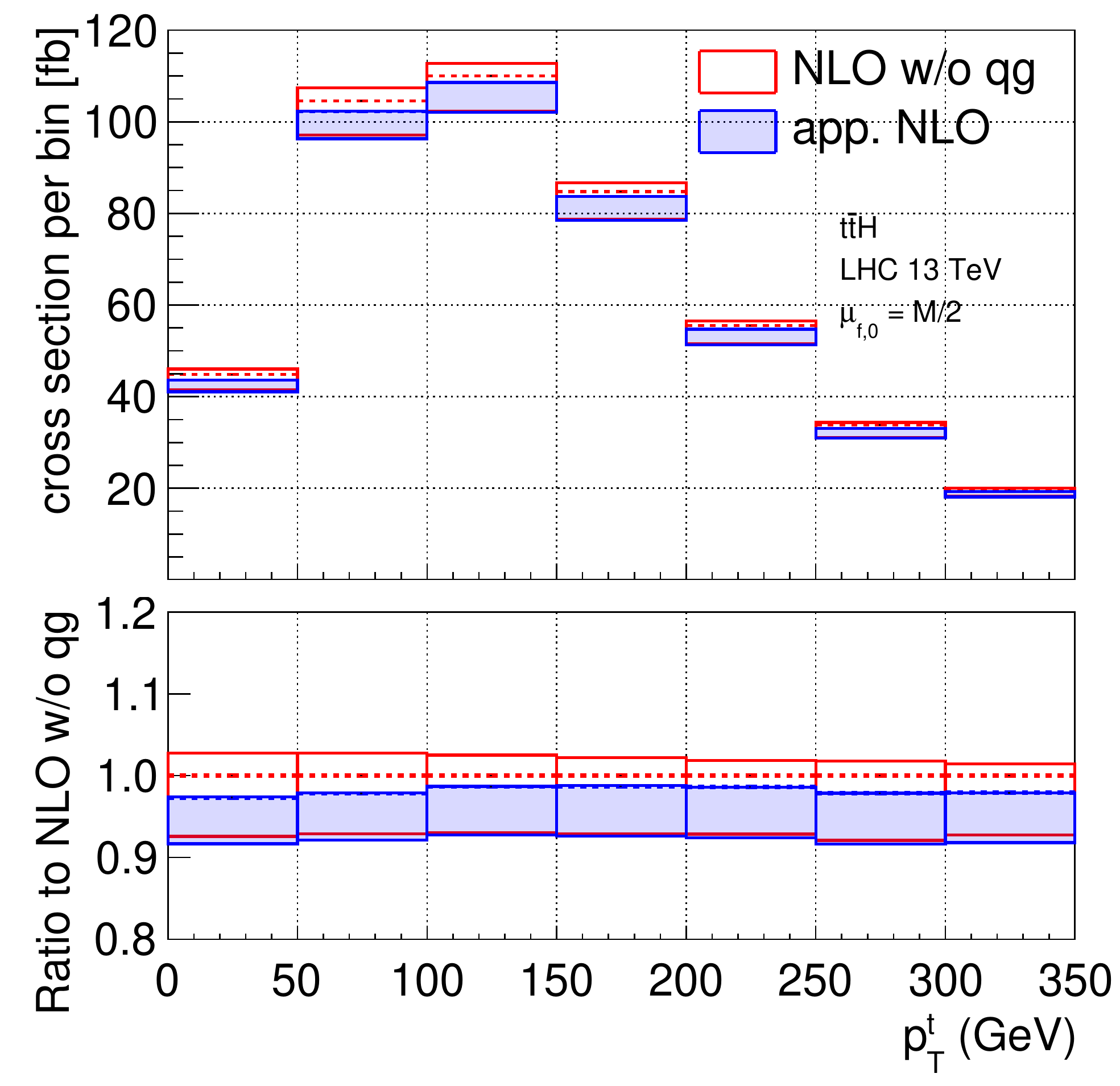} & \includegraphics[width=7.2cm]{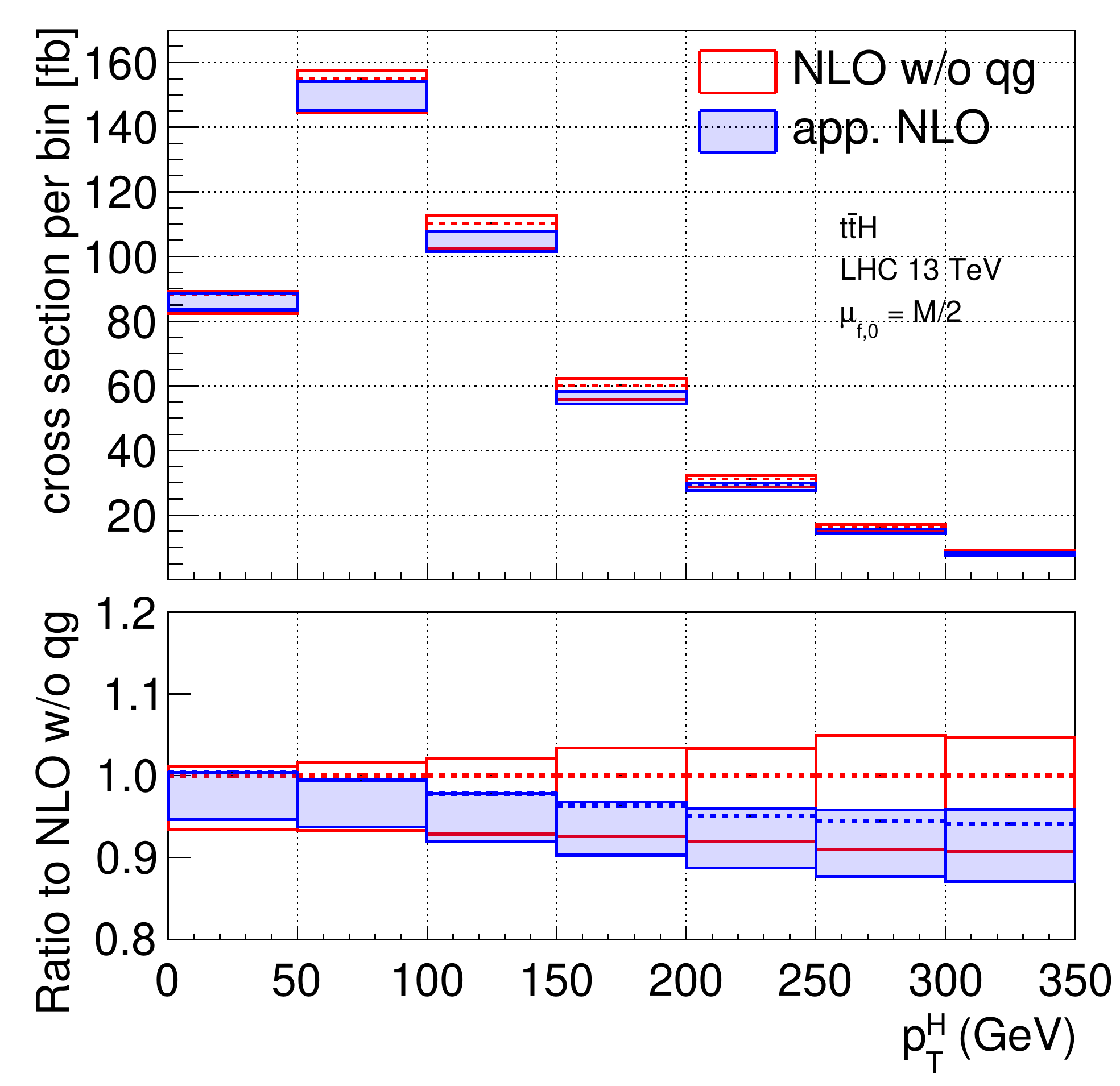} \\
		\end{tabular}
	\end{center}
	\caption{Differential distributions at approximate NLO (blue band) compared to the NLO distributions without the quark-gluon channel contribution (red band). All settings are as in Figure~\ref{fig:nLOvsNLOhalfM}.
		\label{fig:nLOvsNLOnoqghaflM}
	}
	
\end{figure}

As for the case of the total cross section, it is reasonable to look
at how the approximate NLO distributions compare to the NLO
calculations when the contribution of the $qg$ channel is left out
from the latter. This comparison can be found in
Figure~\ref{fig:nLOvsNLOnoqghaflM}. One can see that approximate NLO
and NLO distributions without the $qg$ channel agree quite well and
the size of the respective uncertainty bands is very
similar.
We remind the reader that the contribution of the $qg$-channel at NLO
is included in the NLO+NLL, NLO+NNLL and nNLO predictions discussed
below through the matching procedure.

\begin{figure}[tp]
	\begin{center}
		\begin{tabular}{cc}
			\includegraphics[width=7cm]{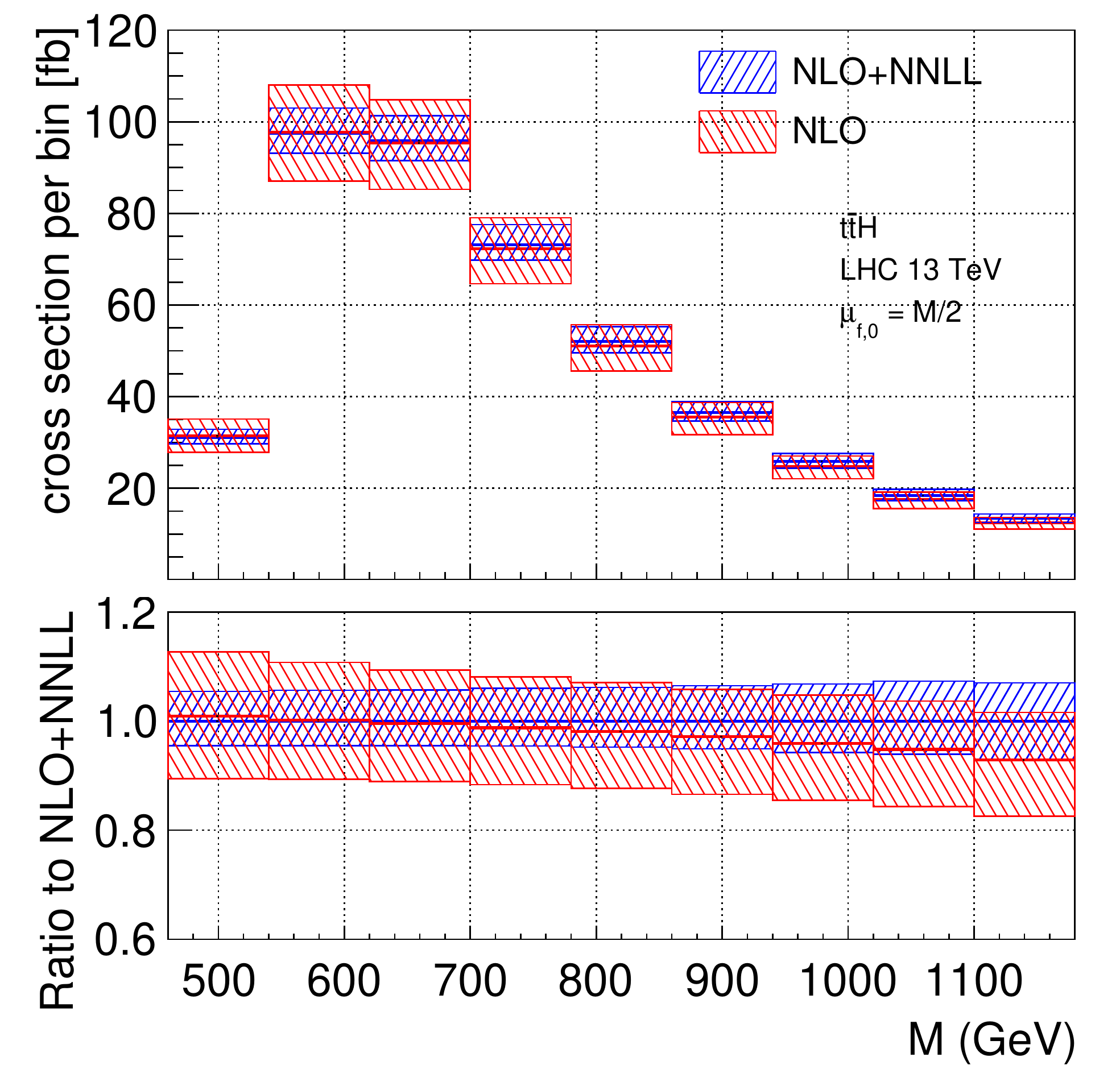} & \includegraphics[width=7cm]{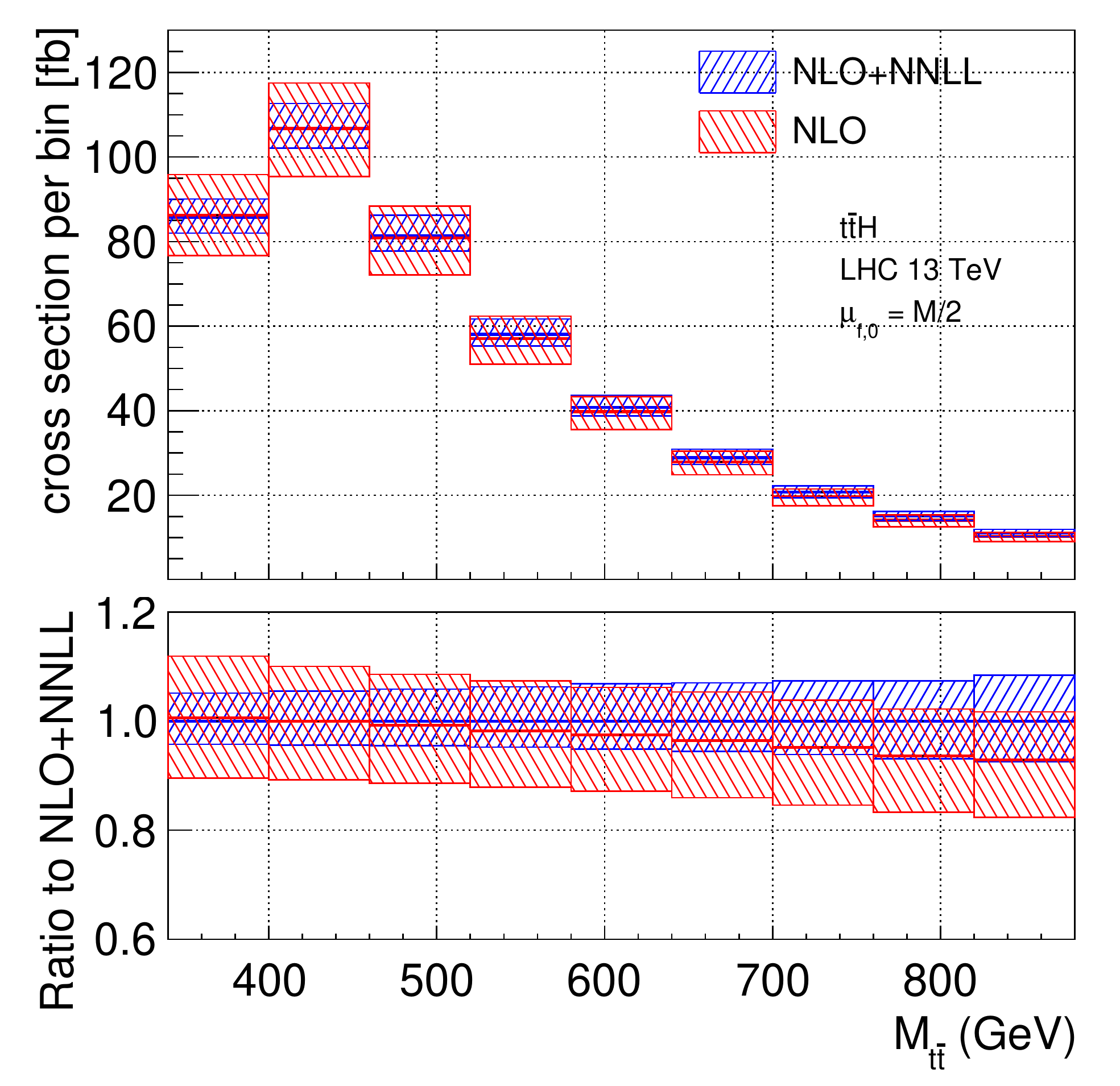} \\
			\includegraphics[width=7cm]{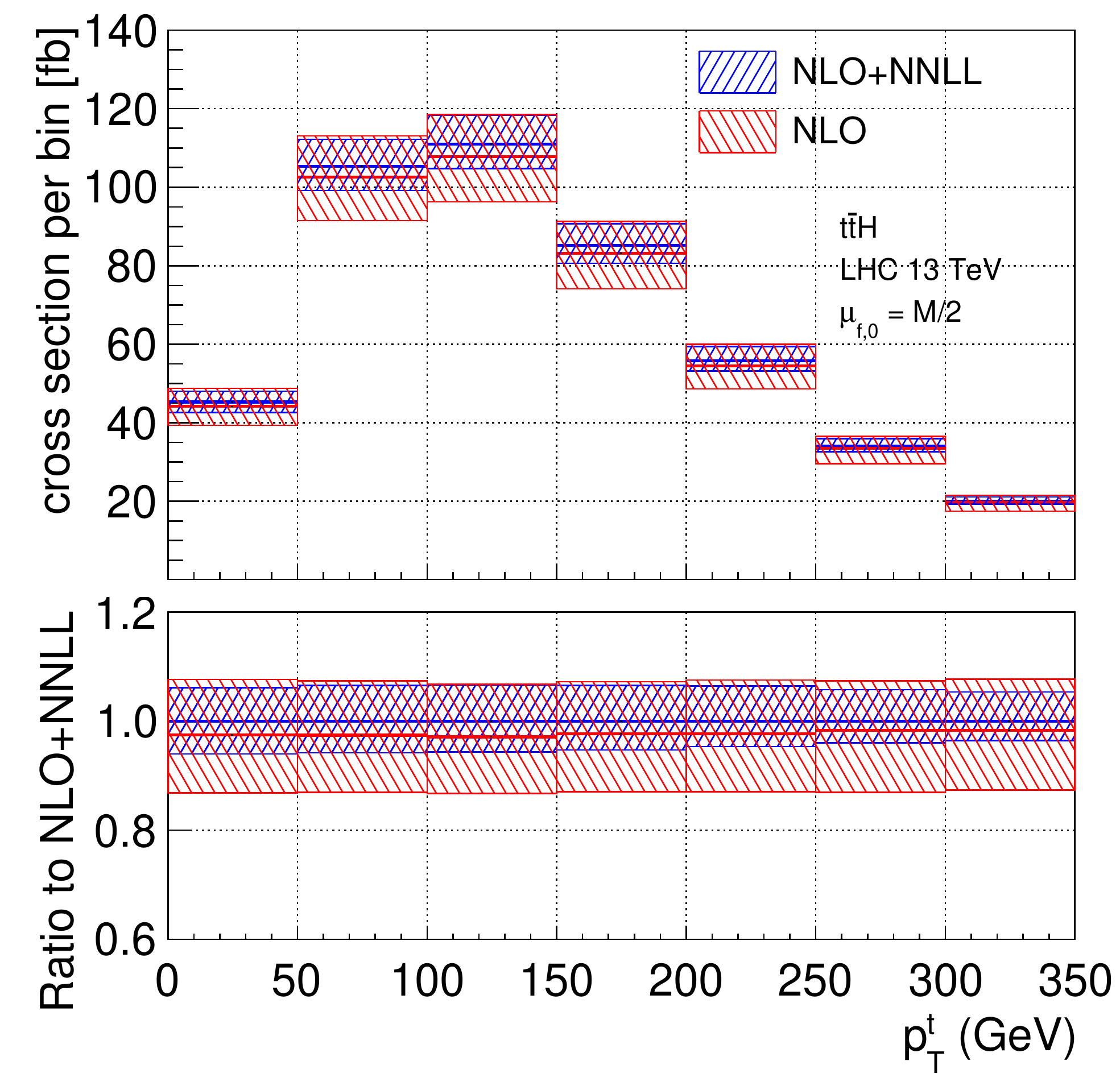} & \includegraphics[width=7cm]{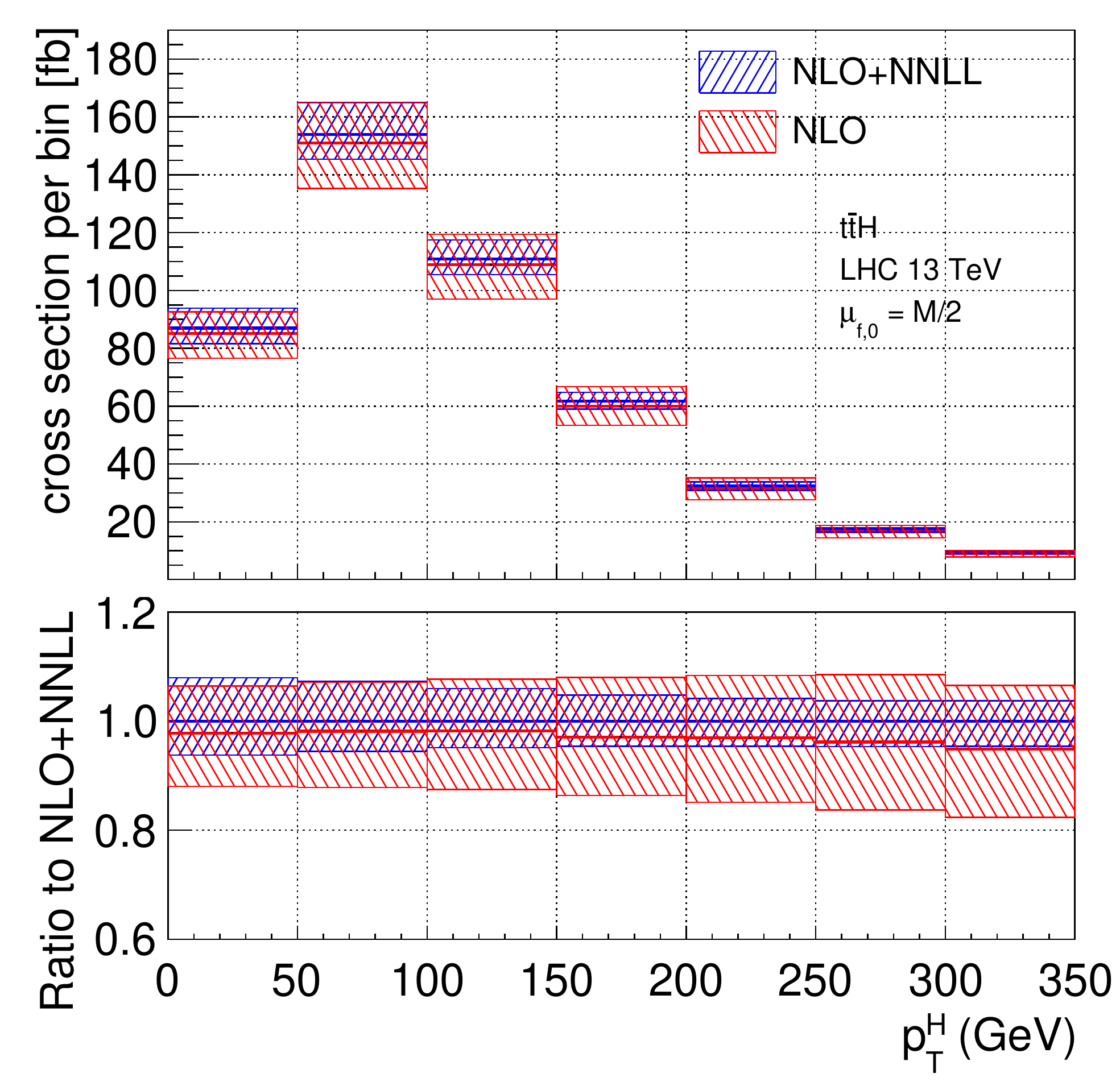} \\
		\end{tabular}
	\end{center}
	\caption{Differential distributions with $\mu_{f,0}=M/2$ at NLO+NNLL (blue band) compared to the NLO calculation (red band). The uncertainty bands
are generated through scale variations of $\mu_f$, $\mu_s$ and $\mu_h$ as 
explained in the text.
			\label{fig:NLOvsNNLLhalfM}
                      }
\end{figure}

The comparison between the NLO and the NLO+NNLL calculations of the
differential distributions can be found in
Figure~\ref{fig:NLOvsNNLLhalfM}. We see that the NLO+NNLL uncertainty
band is included in the NLO scale uncertainty band in almost all bins
of the distributions considered here. The exception is the bins in
the far tail of the $M$ and $M_{t\bar{t}}$ distributions, where the
NLO+NNLL band is not completely included in the NLO one, but is
higher than the NLO one. In general one can observe that the central
value of the NLO+NNLL calculation is slightly larger than the central
value of the NLO one in almost all bins of the distributions shown in
Figure~\ref{fig:NLOvsNNLLhalfM}.

\begin{figure}[tp]
	\begin{center}
		\begin{tabular}{cc}
			\includegraphics[width=7cm]{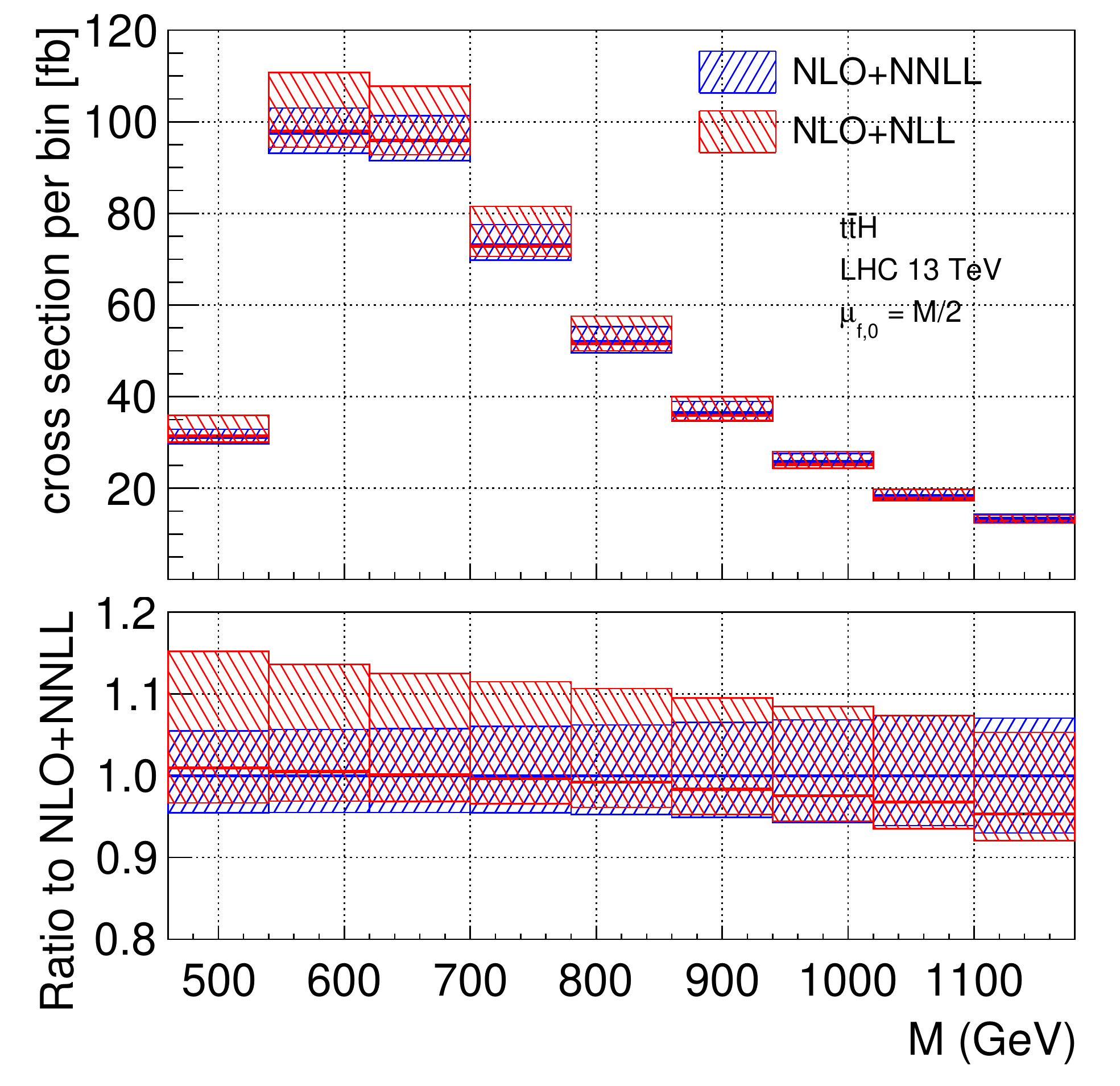} & \includegraphics[width=7cm]{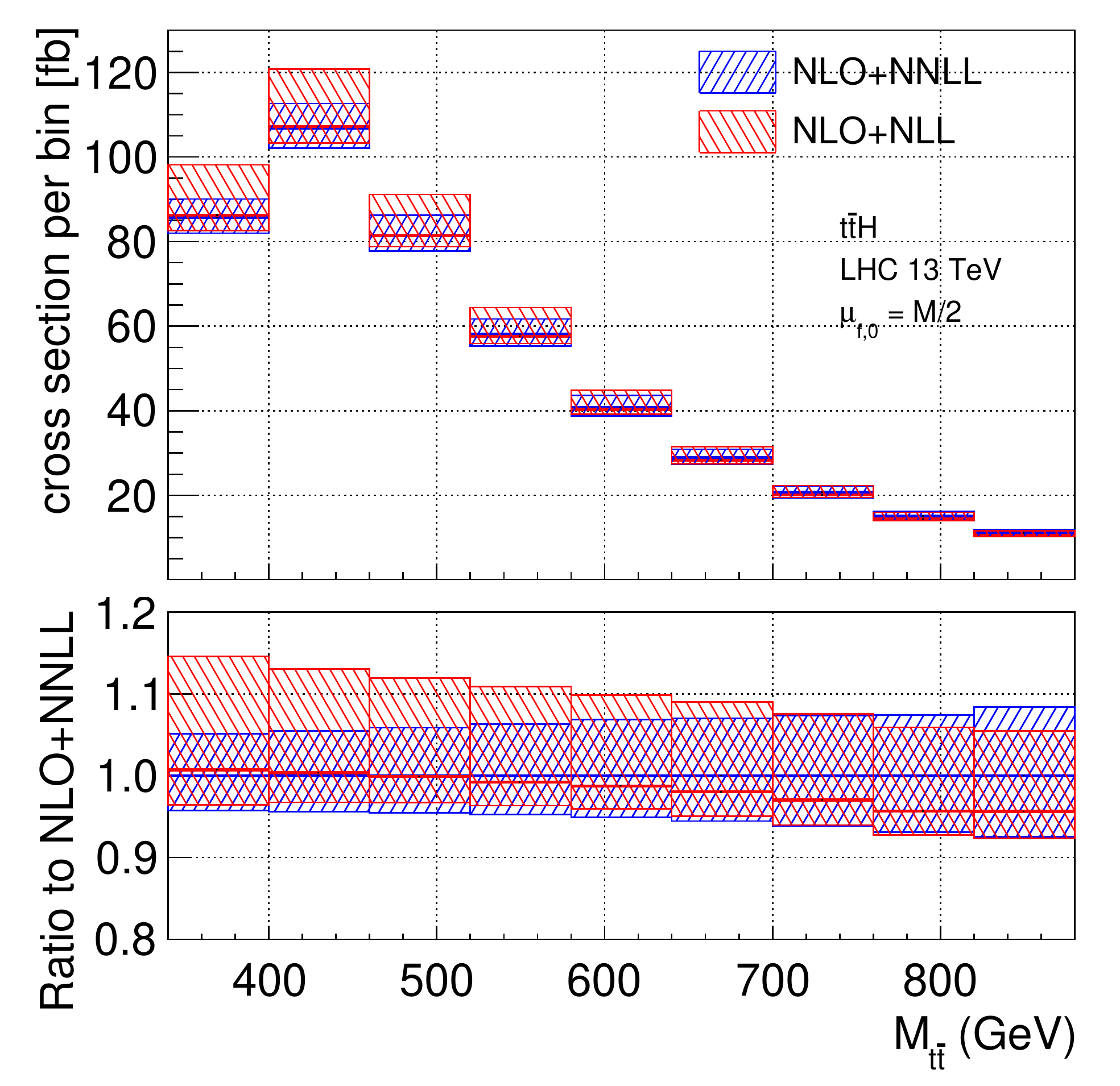} \\
			\includegraphics[width=7cm]{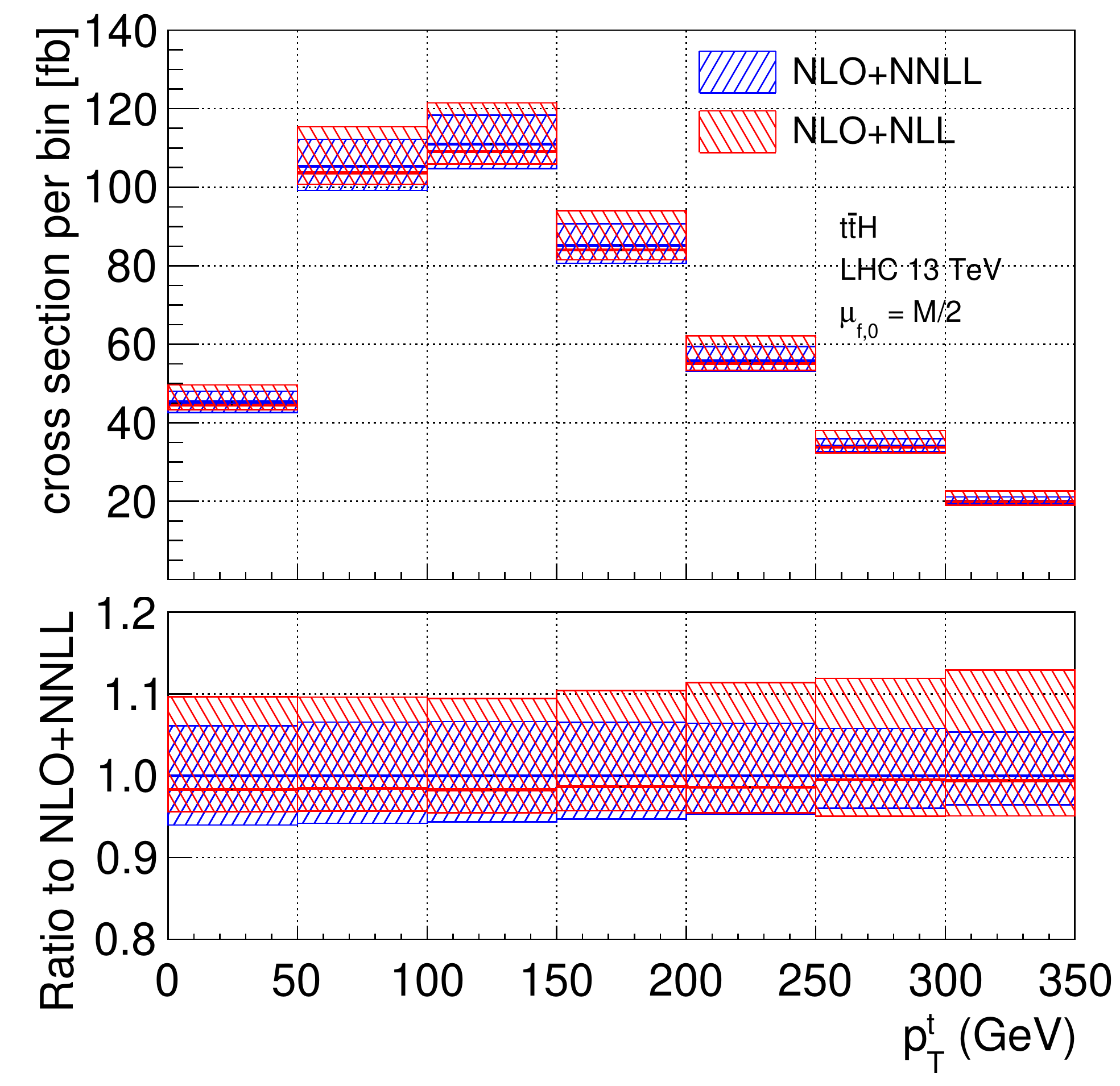} & \includegraphics[width=7cm]{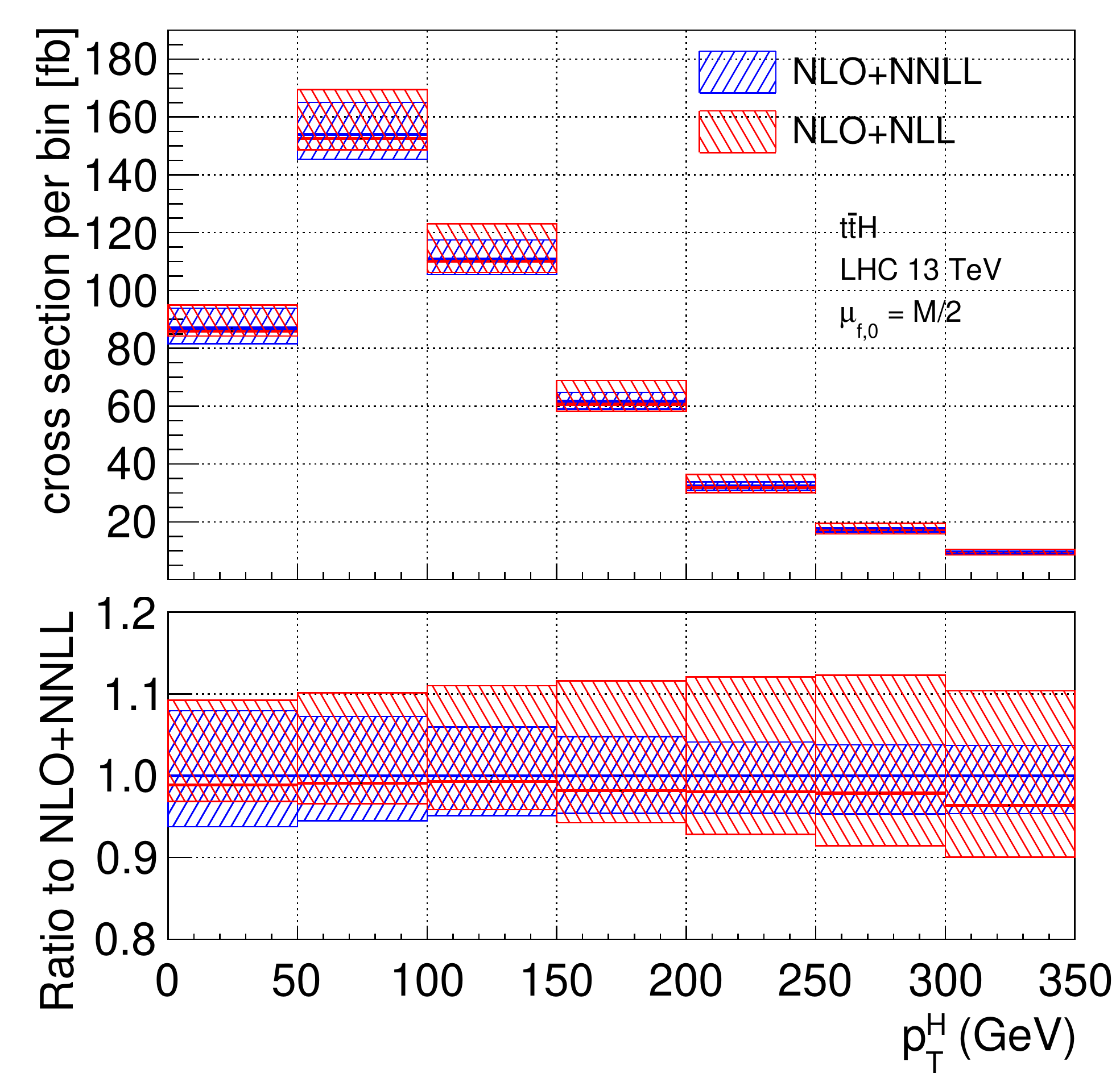} \\
		\end{tabular}
	\end{center}
	\caption{Differential distributions  $\mu_{f,0}=M/2$ at NLO+NNLL (blue band) compared to the NLO+NLL calculation (red band).   The uncertainty bands
are generated through scale variations.
		\label{fig:NLLvsNNLLhalfM}
              }
\end{figure}

Figure~\ref{fig:NLLvsNNLLhalfM} shows a comparison between NLO+NLL and NLO+NNLL results. The central value of these two calculations is quite close in all bins. The main effect of the corrections at NLO+NNLL is to shrink slightly the scale uncertainty bands with respect to the NLO+NLL results everywhere with the exception of the bins in the far tail of the $M$ and $M_{t \bar{t}}$ distributions.

\begin{figure}[tp]
	\begin{center}
		\begin{tabular}{cc}
			\includegraphics[width=7cm]{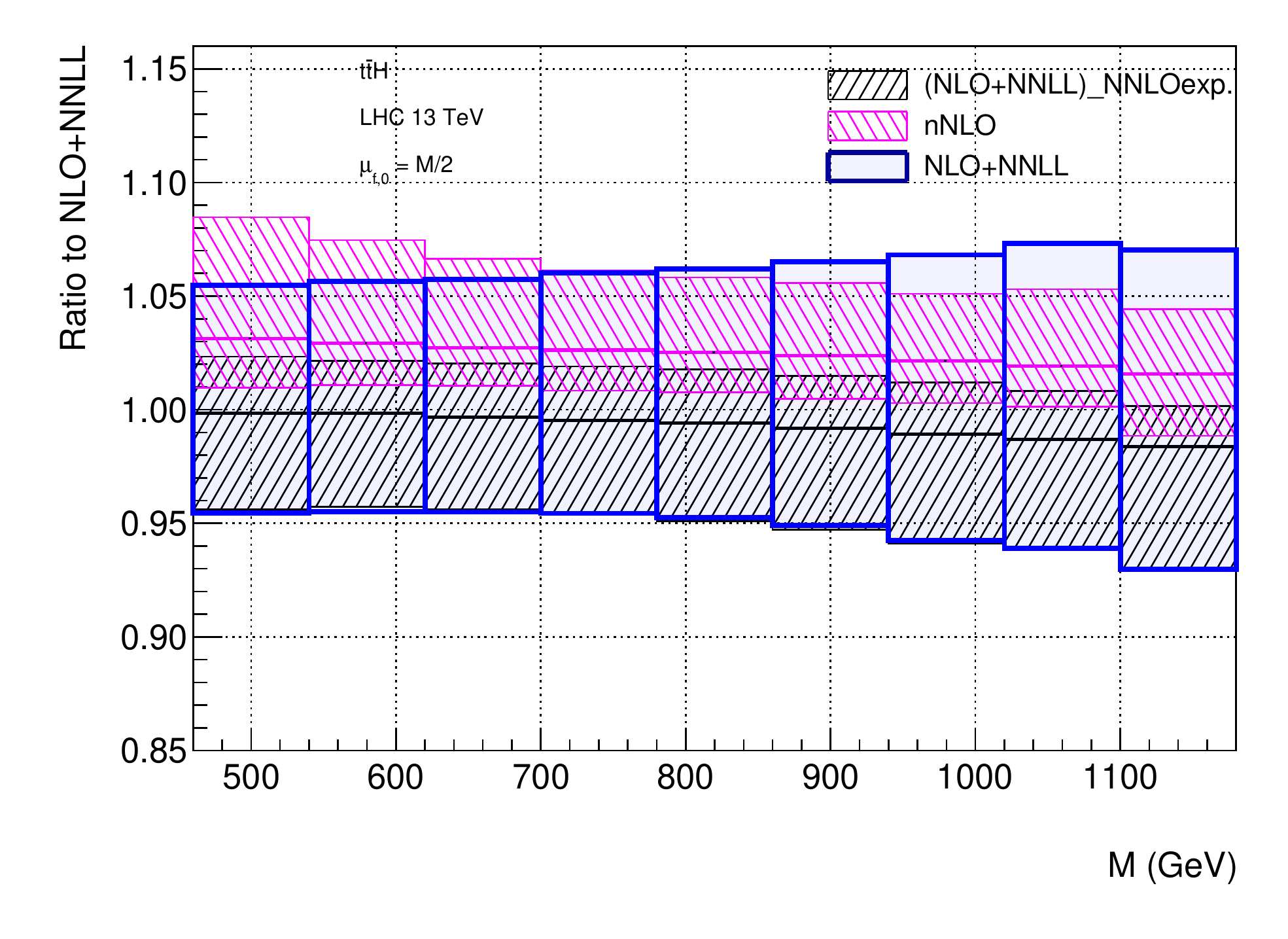} & \includegraphics[width=7cm]{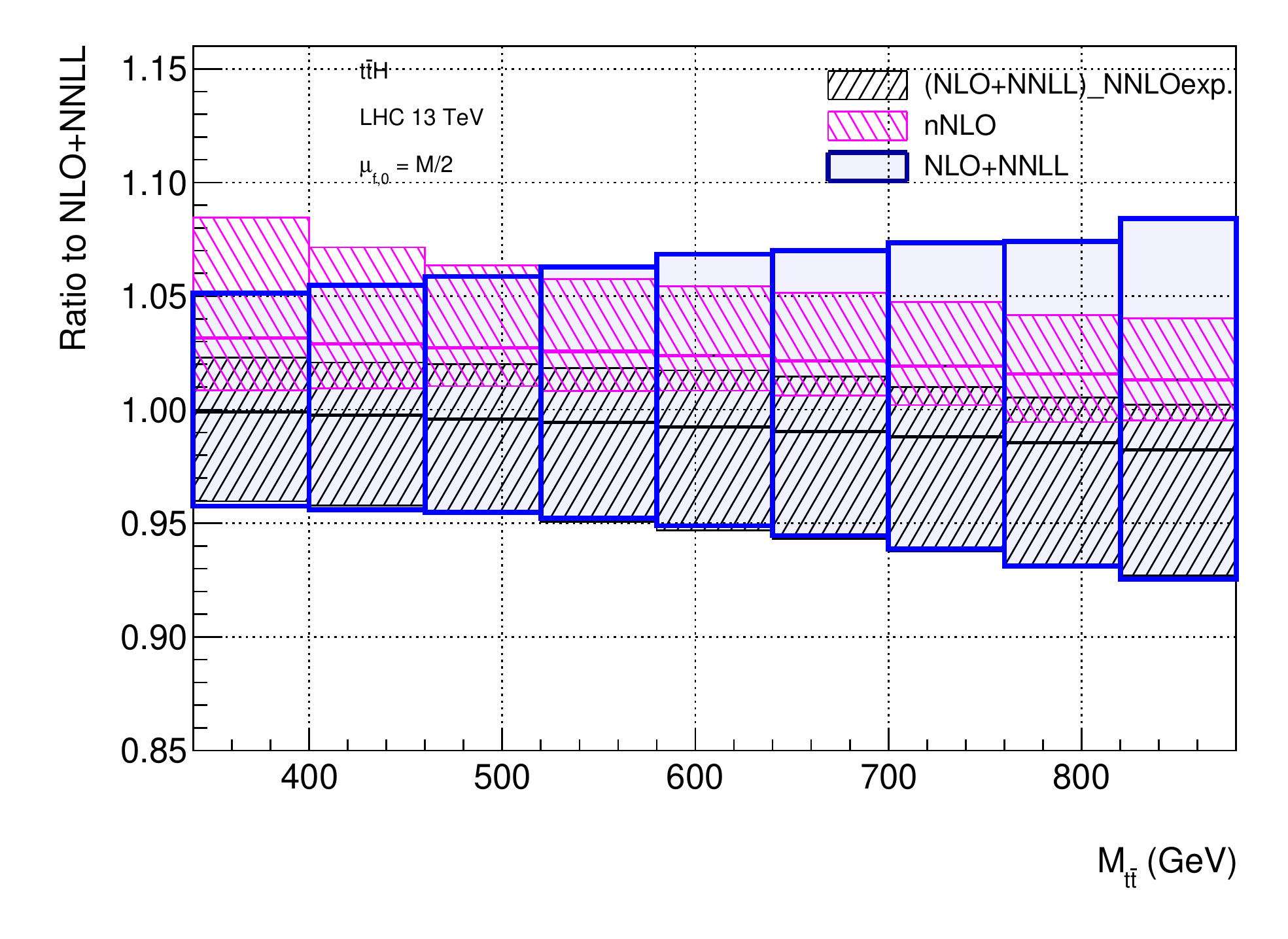} \\
			\includegraphics[width=7cm]{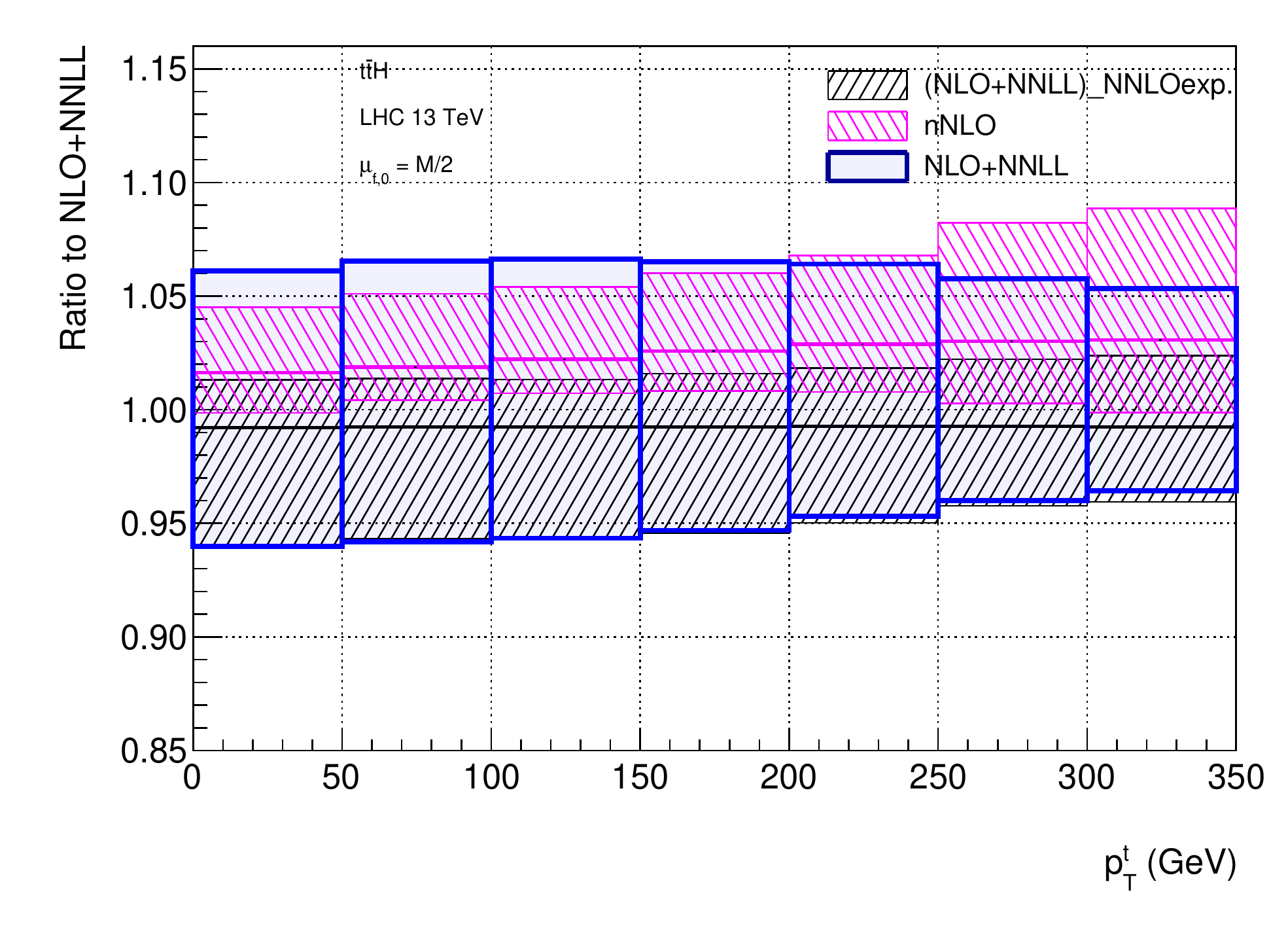} & \includegraphics[width=7cm]{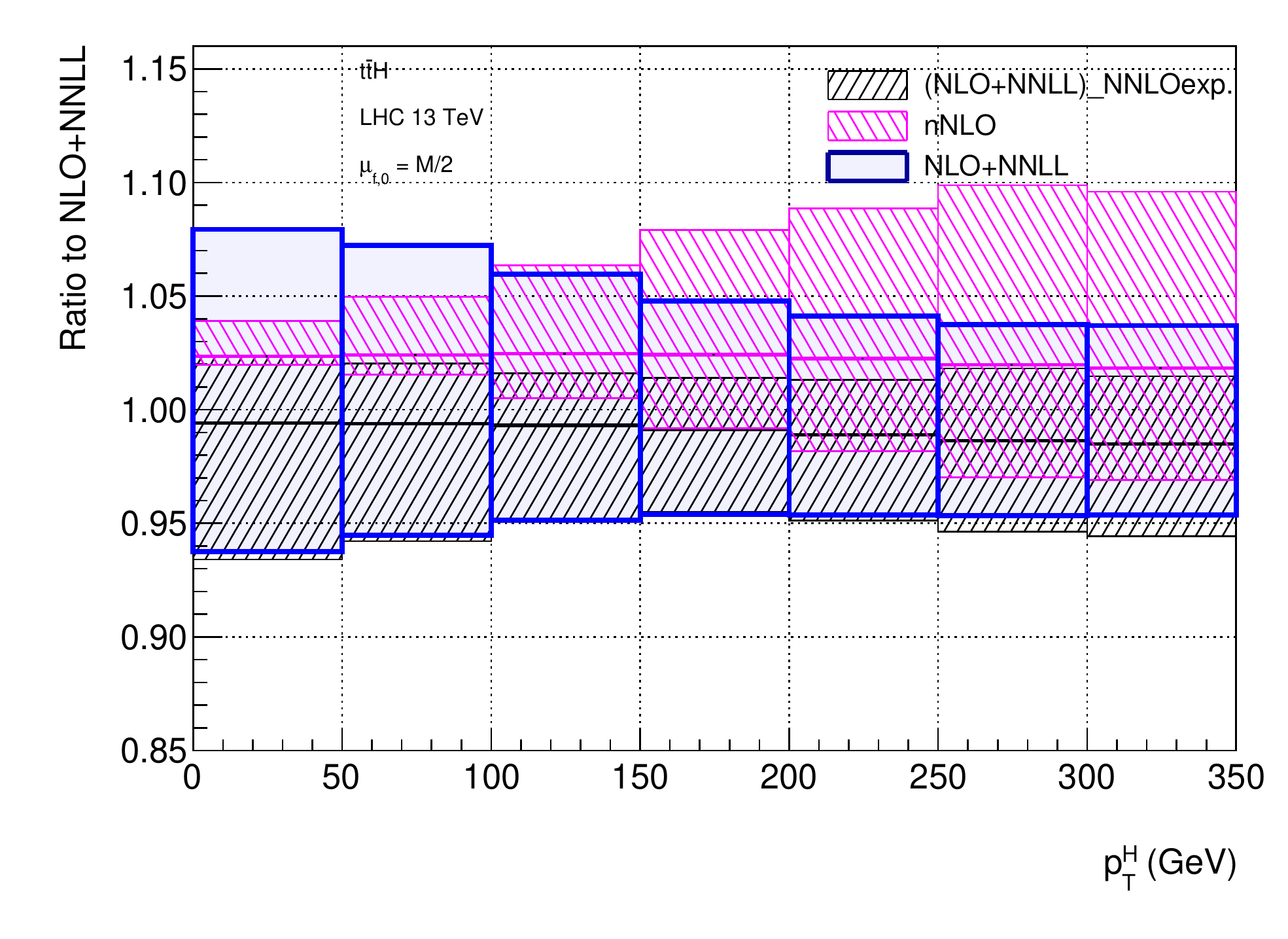} \\
		\end{tabular}
	\end{center}
	\caption{Differential distributions ratios for $\mu_{f,0}=M/2$, where 
the uncertainties are generated through scale variations.
		\label{fig:NNLOratshalfM}
	}
\end{figure}

We conclude our discussion of the results obtained with the choice
$\mu_{f,0} = M/2$ by comparing in Figure~\ref{fig:NNLOratshalfM} the
NLO+NNLL, nNLO and NLO+NNLL expanded predictions for the various
distributions. The figure shows the ratio, separately for each bin, of
the distribution to the NLO+NNLL prediction evaluated with $\mu_i =
\mu_{i,0}$ for $i = s,f,h$. The blue band refers to NLO+NNLL
calculations, the dashed red band to nNLO calculations and the dashed
black band to the NNLO expansion of the NLO+NNLL resummation.  The
  dashed black band and the blue band thus differ by NNLL resummation
  effects of order N$^3$LO and higher.  Numerically, these effects
  contribute roughly at the 5\% level, and as for the total cross
  section the NNLO truncation of the NLO+NNLL resummation formula
  tends to underestimate the uncertainty of the all-orders
  resummation.  The difference between the dashed red band and the
  dashed black band is due to constant NNLO corrections, which are of
  N$^3$LL order.  Taking the envelope of the two NNLO approximations
  (i.e. the black and red bands) gives a more realistic estimate of
  the scale uncertainty, which is generally contained within the
  NLO+NNLL result (the exception is the high-$p_{T}^H$ bins).

\begin{figure}[tp]
	\begin{center}
		\begin{tabular}{cc}
			\includegraphics[width=7cm]{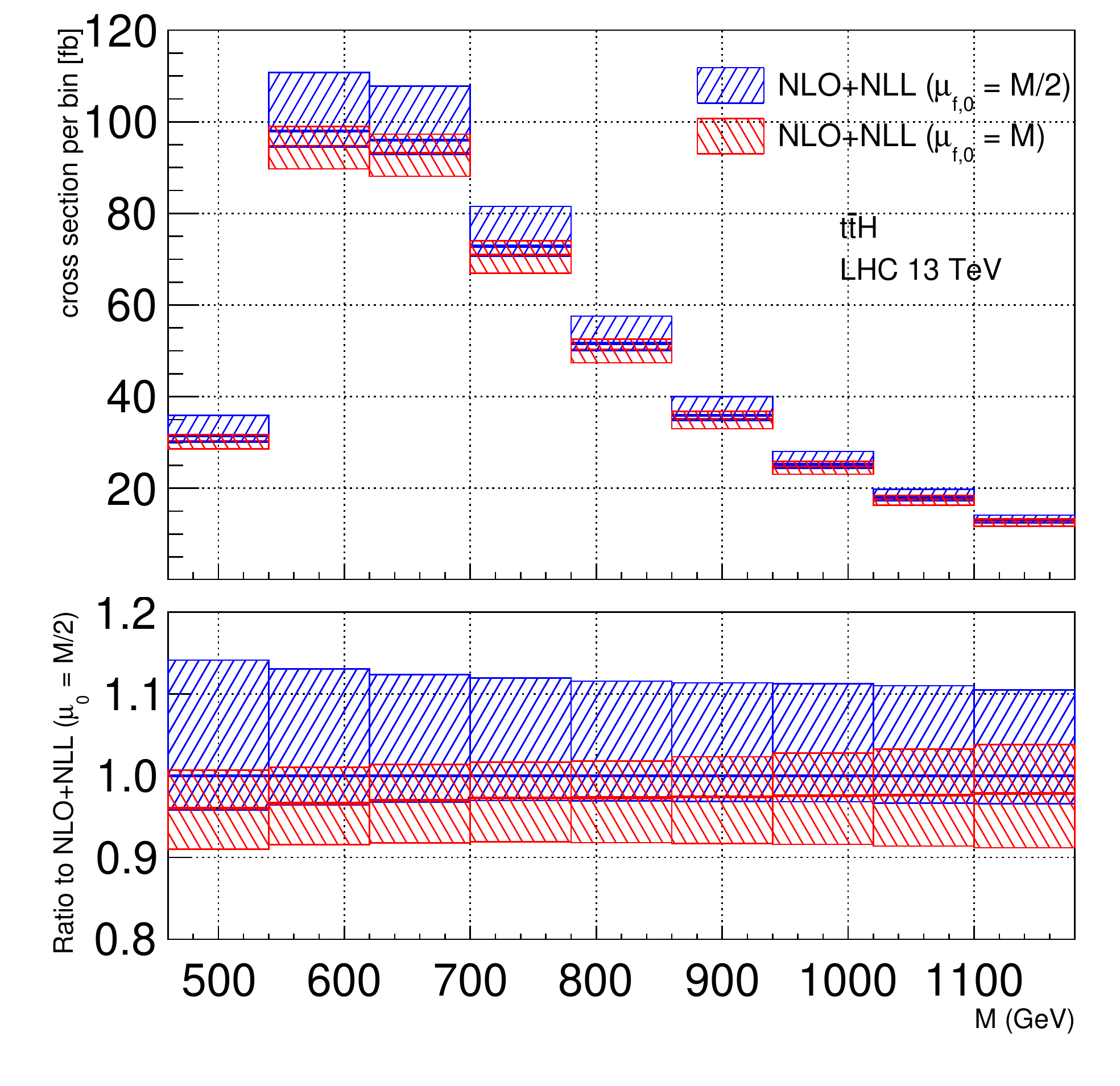} & \includegraphics[width=7cm]{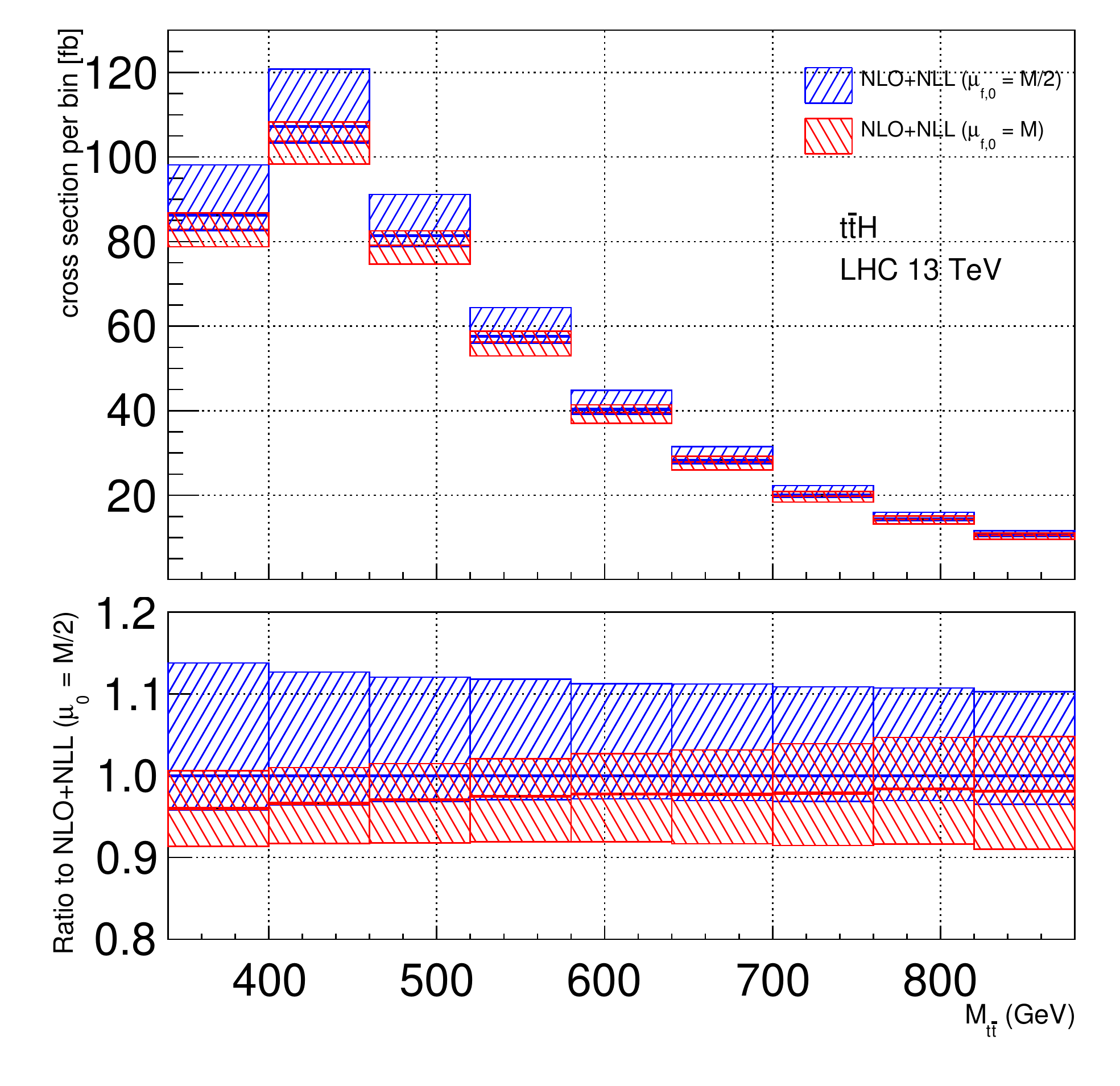} \\
			\includegraphics[width=7cm]{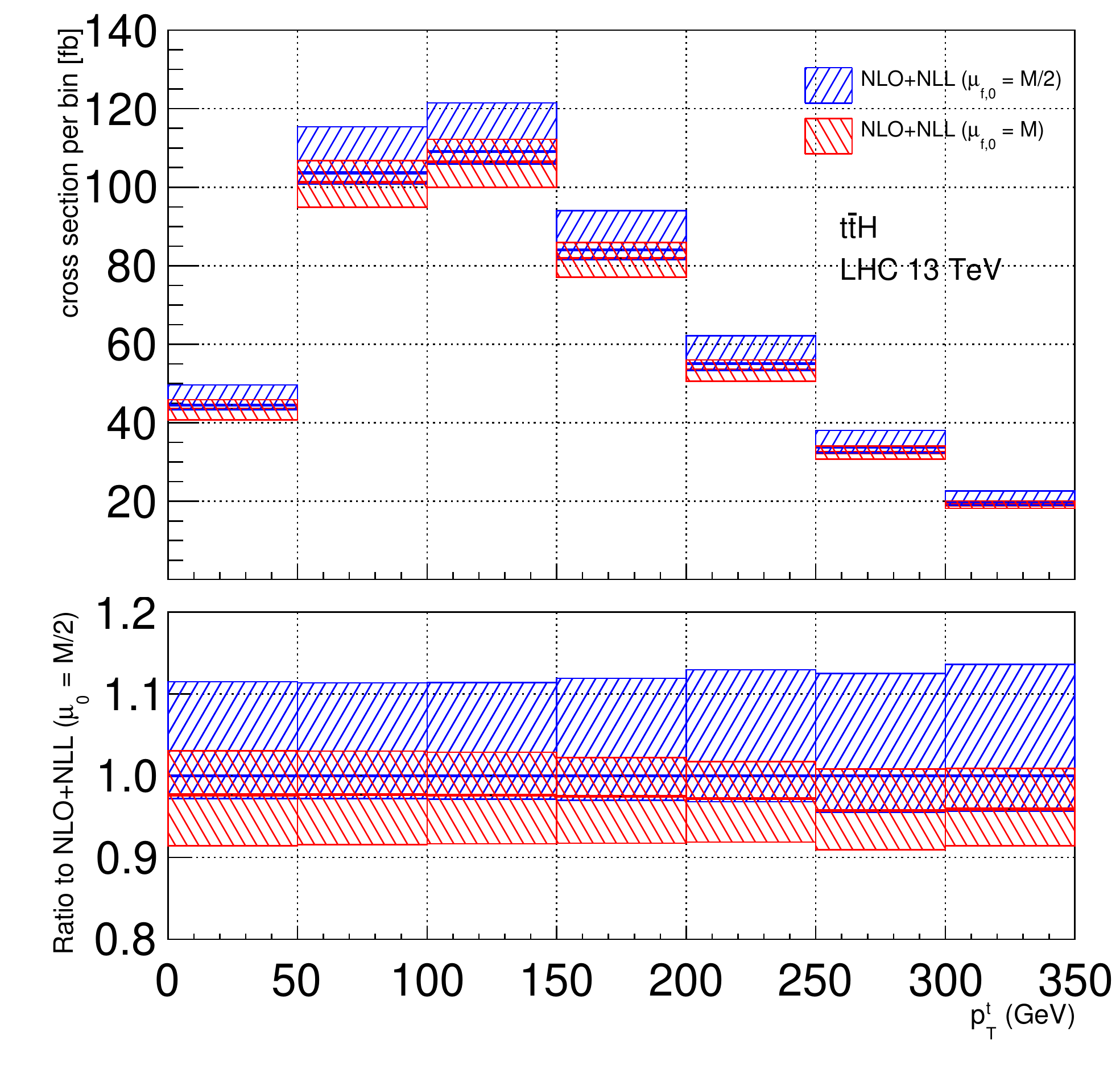} & \includegraphics[width=7cm]{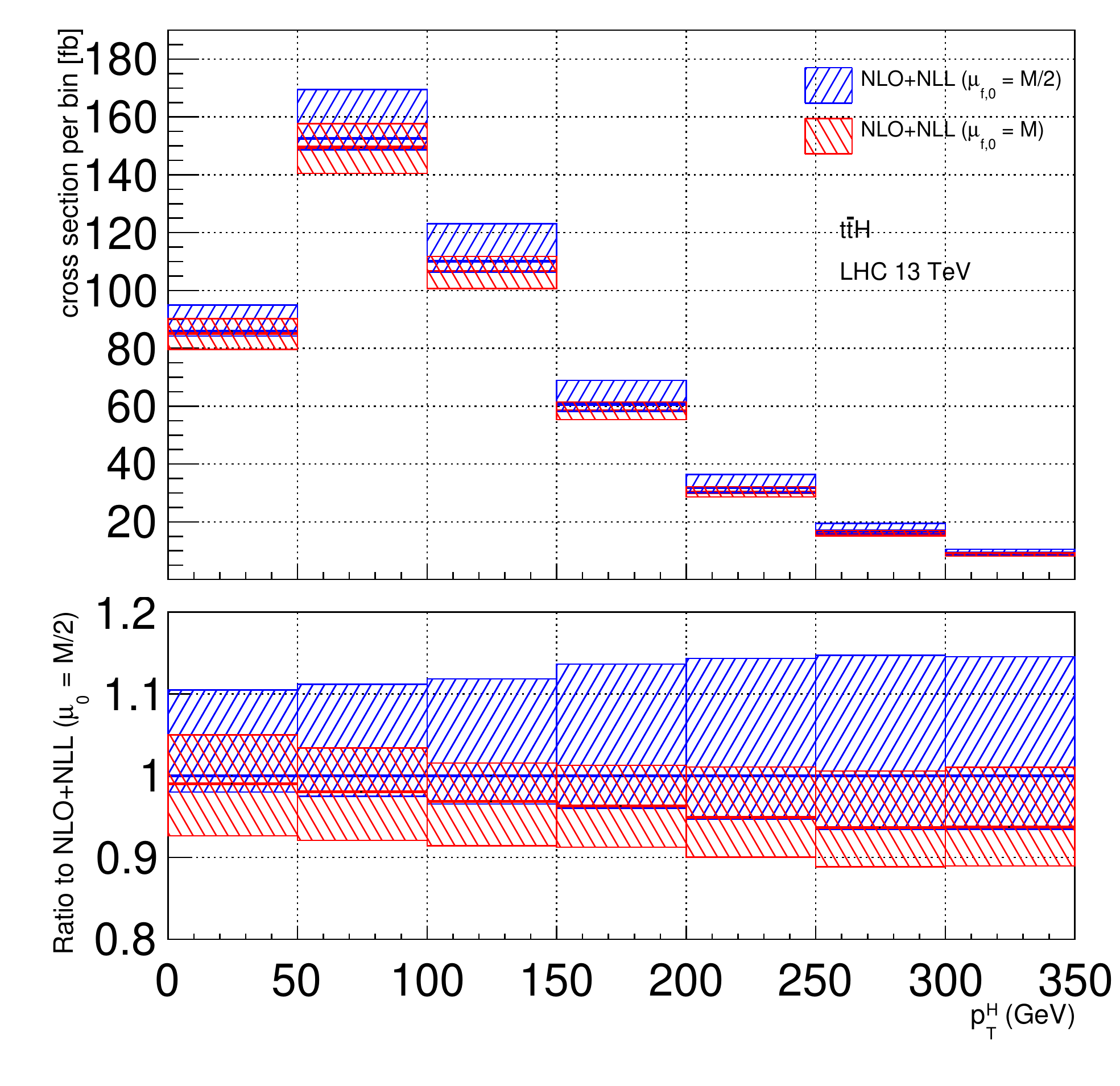} \\
		\end{tabular}
	\end{center}
	\caption{Differential distributions at NLO+NLL at $\mu_{f,0}=M/2$ (blue band) compared to the NLO+NLL calculation at $\mu_{f,0}=M$ (red band), where 
the uncertainties are generated through scale variations.
	%	MMHT 2014 NLO PDFs were used in both cases. 
		\label{fig:NLLvsNLLdiffscales}
	}
\end{figure}

\begin{figure}[tp]
	\begin{center}
		\begin{tabular}{cc}
			\includegraphics[width=7cm]{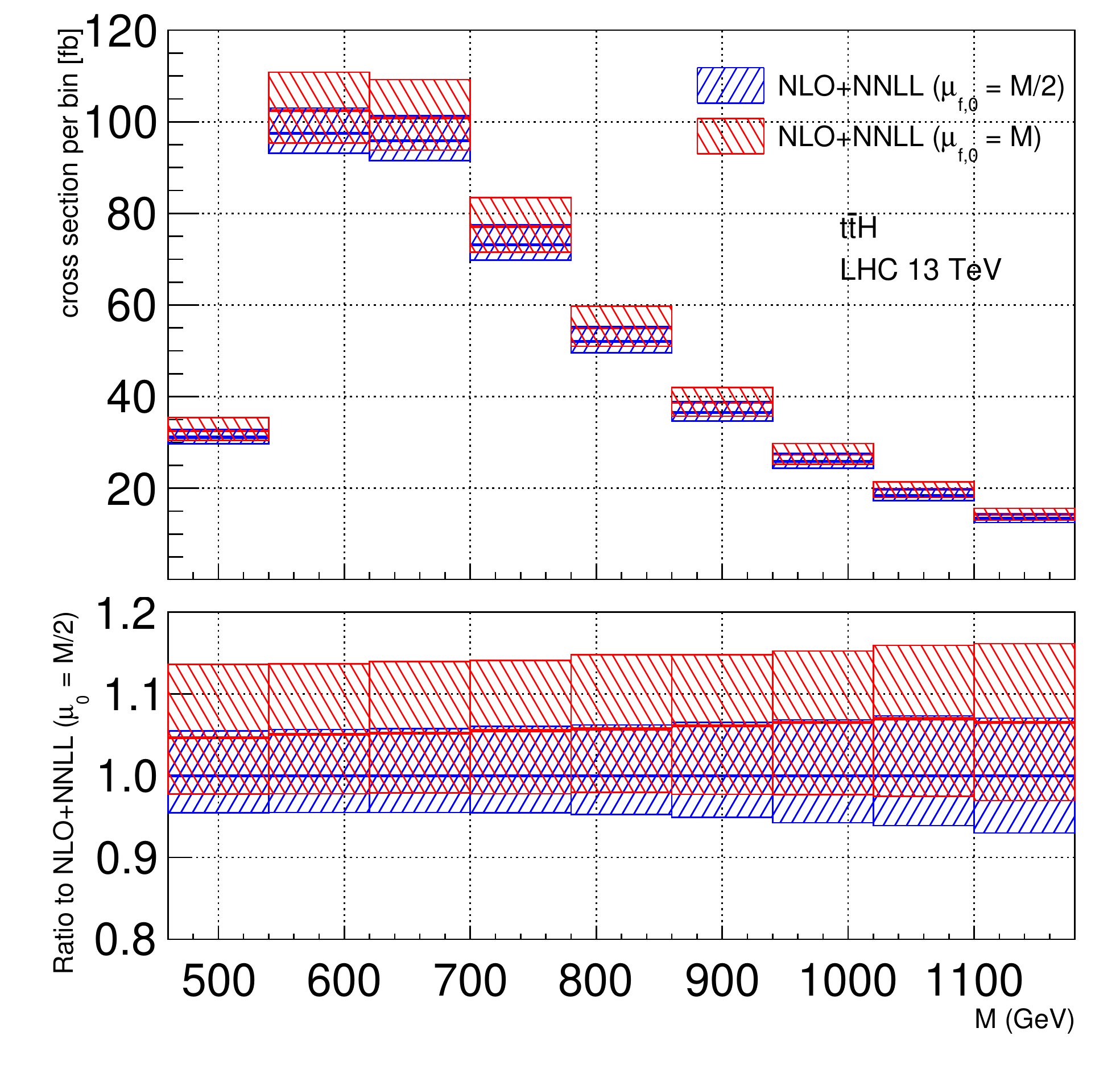} & \includegraphics[width=7cm]{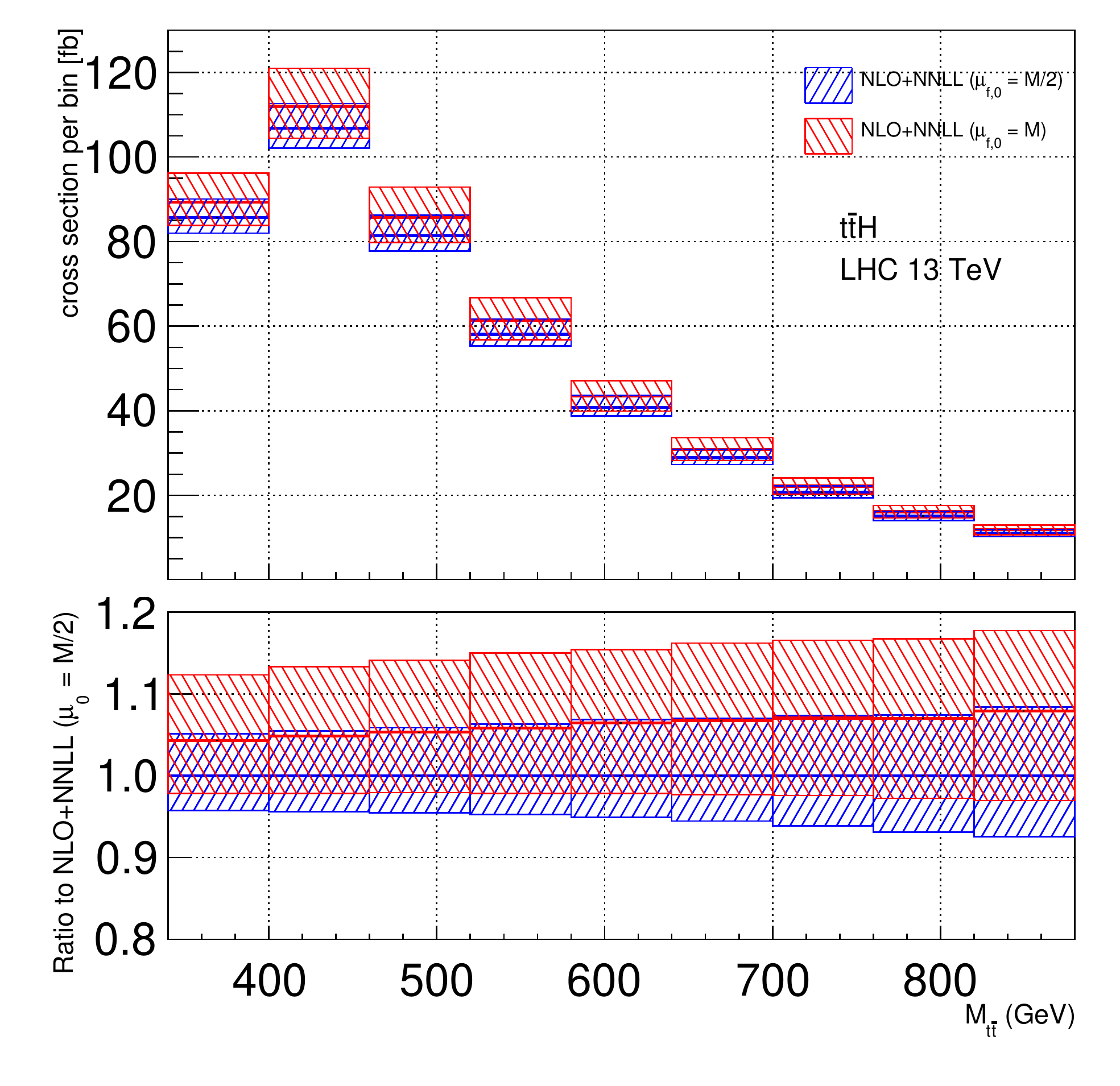} \\
			\includegraphics[width=7cm]{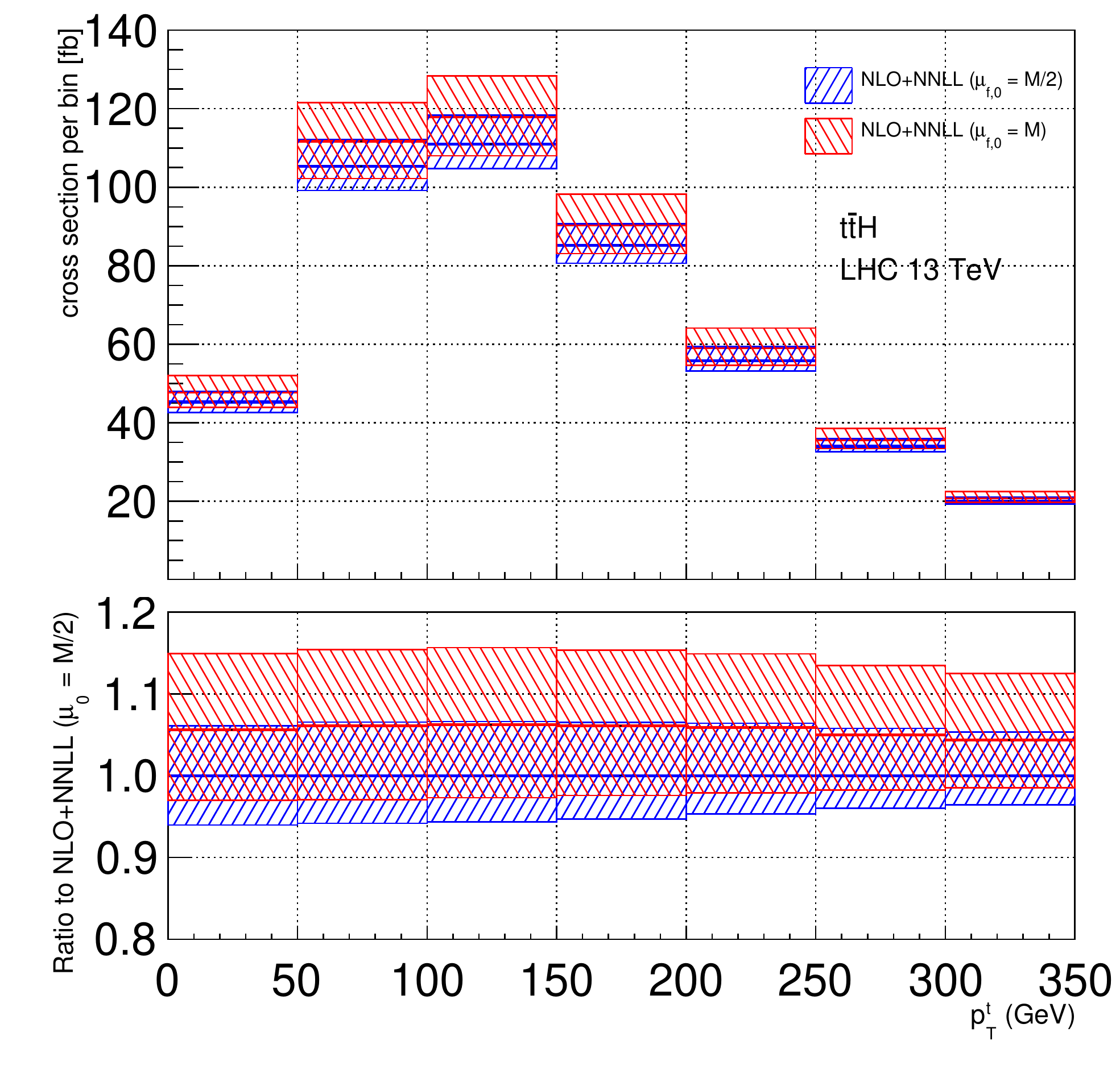} & \includegraphics[width=7cm]{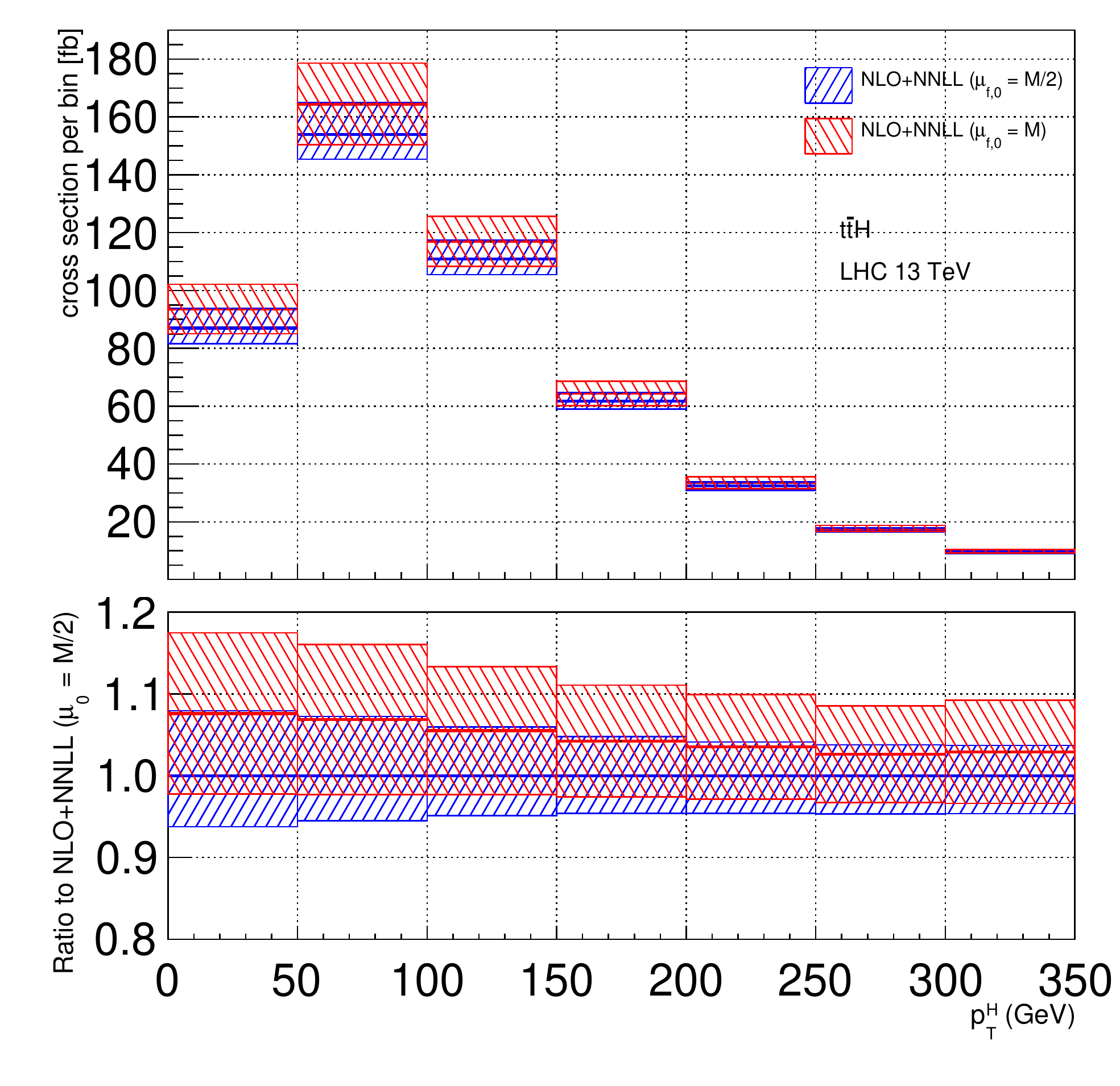} \\
		\end{tabular}
	\end{center}
	\caption{Differential distributions at NLO+NNLL at $\mu_{f,0}=M/2$ (blue band) compared to the NLO+NNLL calculation at $\mu_{f,0}=M$ (red band),  where 
the uncertainties are generated through scale variations.
		\label{fig:NNLLvsNNLLdiffscales}
	}
\end{figure}

We want at this point to study results for a different choice of the
default factorization scale, namely $\mu_{f,0} = M$. As
discussed for the case of the total cross section in Section~\ref{sec:CS}, the
numerical impact of the soft emission corrections with the choice
$\mu_{f,0} = M$ is significantly larger than the impact of the same
corrections with the choice $\mu_{f,0} = M/2$. However, NLO+NNLL
predictions obtained with the two choices are in good agreement.
For what concerns the differential distributions studied here this can
be seen by comparing NLO+NLL calculations carried out with the choice
$\mu_{f,0} = M$ or $\mu_{f,0} = M/2$
(Figure~\ref{fig:NLLvsNLLdiffscales}), and NLO+NNLL calculations with
$\mu_{f,0} = M$ or $\mu_{f,0} = M/2$
(Figure~\ref{fig:NNLLvsNNLLdiffscales}). Figure~\ref{fig:NLLvsNLLdiffscales}
shows that at NLO+NLL the overlap between the distributions evaluated
at $\mu_{f,0} = M$ and $\mu_{f,0} = M/2$ is not particularly good,
with the band at $\mu_{f,0} = M/2$ slightly larger than the one at
$\mu_{f,0} = M$ in all bins.
Figure~\ref{fig:NNLLvsNNLLdiffscales} shows instead that the NLO+NNLL
distributions at $\mu_{f,0} = M$ and $\mu_{f,0} = M/2$ have a large
overlap in all bins. The scale uncertainty at NLO+NNLL with $\mu_{f,0}
= M$ is larger than the scale uncertainty at $\mu_{f,0} = M/2$ in all
bins. The good agreement between the two bands shown in each panel of
Figure~\ref{fig:NNLLvsNNLLdiffscales} indicates that NLO+NNLL
predictions are quite stable with respect to different (but
reasonable) choices of the standard value for the factorization scale.

% % % % % % % % %	

\section{Conclusions}
\label{sec:conclusions}

In this paper we evaluated the resummation of the soft emission
corrections to the associated production of a top-quark pair and a
Higgs boson at the LHC in the partonic threshold limit $z \to 1$. The
calculation is carried out to NNLL accuracy and it is matched to the
complete NLO cross section in QCD. The numerical evaluation of 
observables at NLO+NNLL was carried out by means of
an in-house parton level Monte Carlo code developed for this work, 
based on the resummation formula derived in \cite{Broggio:2015lya}. 
The resummation procedure is however carried out in Mellin space, following the same
approach employed in \cite{Ferroglia:2015ivv,Pecjak:2016nee} for the
calculation of the (boosted) top-quark pair production cross section
and in \cite{Broggio:2016zgg} for the calculation of the cross section
for the associated production of a top-quark pair and a $W$ boson.

In the previous sections we presented predictions for the total cross section 
for this production process at the LHC operating at a center-of-mass energy of
$13$~TeV. In addition, we showed results for four different
differential distributions depending on the four-momenta of the
massive particles in the final state: the differential distributions
in the invariant mass of the $t \bar{t} H$ particles, in the invariant mass of the $t
\bar{t}$ pair, in the transverse momentum of the Higgs boson, 
and in the transverse momentum of the top quark.
We found that the relative size of the NNLL corrections with respect to
the NLO cross section is rather sensitive to the choice of the
factorization scale $\mu_f$. In particular, for the two choices which
we analyzed in detail, namely $\mu_{f,0} = M/2$ and $\mu_{f,0} = M$,
it was found that the NNLL corrections enhance the total cross section
and differential distributions in all bins considered. The NNLL soft
emission corrections expressed as a percentage of the NLO observables
are larger at $\mu_{f,0} = M$ than they are at $\mu_{f,0} =
M/2$. However, by comparing NLO+NNLL predictions obtained by setting
$\mu_{f,0} = M/2$ with NLO+NNLL predictions evaluated with $\mu_{f,0} = M$, and after
accounting for the scale uncertainty affecting both predictions, we
find compatible results.   This fact shows that the NLO+NNLL predictions
are quite stable with respect to the factorization scale choice.
Indeed, it would not be unreasonable to combine the envelope of the results
	at the two different scale choices into a single result with a larger perturbative 
	uncertainty, which for the case of the total cross section would be at about the 
	20~\% level. By taking the envelope of the corresponding NLO results, one finds instead an uncertainty larger than 30~\%.
We also studied the total cross section and differential distributions
at NLO+NLL accuracy and with NNLO approximations of the NLO+NNLL resummation formula, and found
that both of these are a poor proxy for the more complete NLO+NNLL results,
especially for higher values of $\mu_{f,0}$.

The parton level Monte Carlo developed for this paper
could be extended to include the decays of the top quarks and the Higgs boson
following the work done in \cite{Broggio:2014yca}. This would allow one to impose cuts on the momenta of the detected particles.
Furthermore, our code could
serve as a template for the calculation of the NNLL soft emission
corrections to the associated production of a top pair and a $Z$ boson
at the LHC. The latter is a process of significant phenomenological
interest which has already been investigated experimentally at both
the Run I and Run II of the LHC. We plan to study the NLO+NNLL cross
section for this process in future work.

\section*{Acknowledgments}
The in-house Monte Carlo code which we developed and employed to
evaluate the (differential) cross sections presented in this paper was run on the computer
cluster of the Center for Theoretical Physics at the Physics Department of New
York City College of Technology.
We thank A. Signer for discussions about the Monte Carlo implementation,  P.~Maierhofer  and 
S.~Pozzorini for their assistance with the program {\tt Openloops}, and  and G.~Ossola for useful discussions.  The work of A.F. is supported in part by the National Science Foundation under
Grant No. PHY-1417354.  B.P. would like to thank the ESI Vienna for hospitality
and support during the completion of this work. The work of L.L.Y. was supported in part by the National Natural
Science Foundation of China under Grant No. 11575004.

\bibliography{mybib}

\bibliographystyle{JHEP}

\end{document}